\pdfoutput=1
\documentclass[aip,jcp,floatfix,reprint,twocolumn]{revtex4-1}

\usepackage{graphicx}
\usepackage{dcolumn}
\usepackage{bm}
\usepackage{color}
\usepackage{multirow}
\usepackage{amsfonts}
\newcommand*{\plimsoll}{{\ensuremath{-\kern-4pt{\ominus}\kern-4pt-}}}

\newcommand{\hilightg}[1]{\colorbox{green}{#1}}
\newcommand{\hilighty}[1]{\colorbox{yellow}{#1}}
\newcommand{\hilightb}[1]{\colorbox{blue}{#1}}

\usepackage{xr}
\usepackage{color}
\newcommand{\teo}[1]{\textcolor{black}{#1}} 

\begin{document}

\title{DNA nanotechnology: understanding and optimisation through simulation}

\author{Thomas E. Ouldridge}

\address{$^1$Rudolf Peierls Centre for Theoretical Physics, University of Oxford Physics Department, 1 Keble Road, Oxford, OX1 3PG, UK, and Department of Mathematics, South Kensington Campus, Imperial College London, London,
SW7 2AZ, UK.}

\begin{abstract}
DNA nanotechnology promises to provide controllable self-assembly on the nanoscale, allowing for the design of static structures, dynamic machines and computational architectures. In this article I review the state-of-the art of DNA nanotechnology, highlighting the need for a more detailed understanding of the key processes, both in terms of theoretical modelling and experimental characterisation. I then consider coarse-grained models of DNA, mesoscale descriptions that have the potential to provide great insight into the operation of DNA nanotechnology if they are well designed. In particular, I discuss a number of nanotechnological systems that have been studied with oxDNA, a recently developed coarse-grained model, highlighting the subtle interplay of kinetic, thermodynamic and mechanical factors that can determine behaviour. Finally, new results highlighting the importance of mechanical tension in the operation of a two-footed walker are presented, demonstrating that recovery from an unintended `overstepped' configuration can be accelerated by three to four orders of magnitude by application of a moderate tension to the walker's track. More generally, the walker illustrates the possibility of biasing strand-displacement processes to affect the overall rate.\\
\emph{Key words}:
DNA nanotechnology; self-assembly; molecular machines; non-equilibrium systems;   coarse-grained modelling; simulation. \\
{*Correspondence: t.ouldridge@imperial.ac.uk\\}
\end{abstract}

\maketitle

\section{Introduction to DNA nanotechnology}
\subsection{The DNA molecule and its potential}
Deoxyribonucleic acid (DNA) is a macromolecule with a backbone of covalently linked sugar and phosphate groups; attached to each sugar is a base, which can be adenine (A), guanine (G), cytosine (C) or thymine (T) \cite{Saenger1984}. DNA is often found as a double helix of antiparallel strands stabilised by hydrogen-bonding of complementary Watson-Crick base pairs  (AT and CG) and stacking interactions between the planar bases. Double-stranded DNA (dsDNA) is stiff, with a persistence length of  $\sim150$ base pairs \cite{Hagerman1988}, whereas unpaired single strands (ssDNA) are far more flexible \cite{Mills1999,Murphy2004}. 

DNA carries information through its sequence of bases - in biology, this information codes for proteins and their regulation. The specificity of Watson-Crick base-pairing means that both strands within a duplex carry identical information, facilitating replication. In 1982, Seeman speculated that the specificity of DNA hybridization could be harnessed to permit the design of artificial structures, proposing that certain sequences could self-assemble into crystals \cite{Seeman1982}.

\subsection{DNA nanostructures}
The Seeman group quickly  constructed artificial nucleic acid junctions \cite{Kallenbach83}, but the first 3-dimensional crystal based purely on rationally designed Watson-Crick base pairs was not demonstrated until 2009 \cite{Zheng2009} (crystals involving non-canonical interactions were created in 2004 \cite{Paukstelis2004}). In the intervening period,  ribbons \cite{Yan2003} and 2D lattices \cite{Winfree98,Malo2005} were realised. Complex Archimedean tiling patterns \cite{Zhang2013} and `empty liquids' \cite{Biffi2013} have recently been achieved through hybridization-driven DNA self-assembly. 

As well as macroscopic phases, DNA can form structures of well-defined finite size. Early successes (cubes \cite{Chen91} and octahedra \cite{Zhang94}) involved several discrete stages of assembly, but it was subsequently shown that polyhedra \cite{Goodman2005, Erben07,Andersen08} and then more complex structures \cite{Wei2012,Ke2012} could be made to form simply  by cooling solutions of short ssDNA strands (oligonucleotides). An alternative approach, known as DNA `origami' \cite{Rothemund06}, uses one long `scaffold' strand that is shaped by shorter `staple' strands into a complex structure. This technique can produce three dimensional objects \cite{Douglas09, Andersen09} and structures with curvature or twist \cite{Dietz2009,Han2011}. Some examples of DNA nanostructures are shown in Fig.\,\ref{nanotech_examples}\,(a)-(c).

DNA does not have to be used in isolation -- it can also be conjugated with other molecules or nanoparticles. Crystals of  DNA-coated colloids have been constructed \cite{Park2008}, and small organic molecules have been used as vertices in structures held together by DNA \cite{Aldaye2007,Zimmermann08}. 

\subsection{Dynamic DNA devices}
\begin{figure*}
\centerline{\includegraphics[width=30pc]{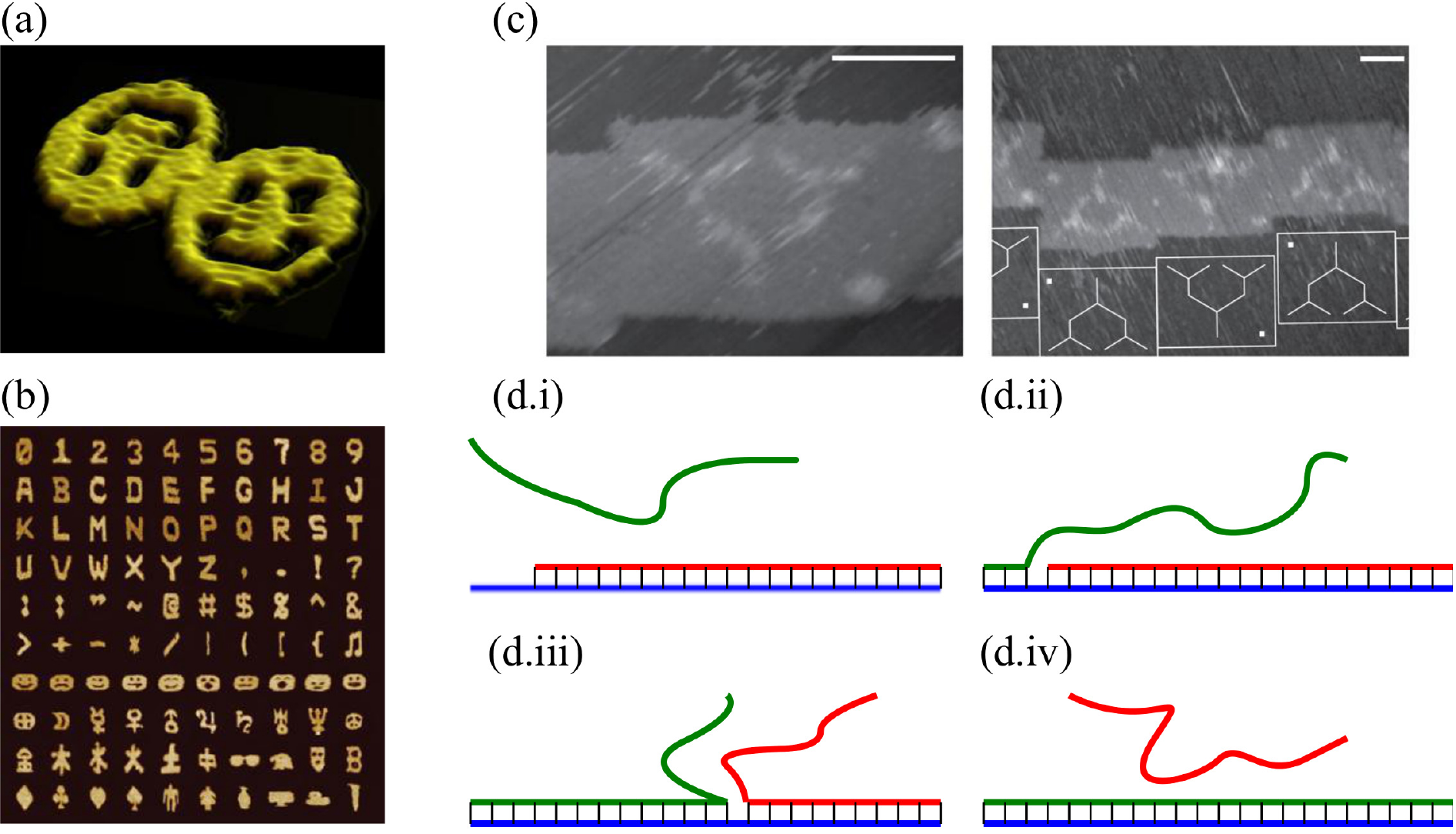}}
\caption{(a)-(c) Examples of DNA nanotechnology. (a) Atomic force microscopy (AFM) image of the original DNA `origami' \cite{Rothemund06}, courtesy of P. W. K. Rothemund. (b) AFM of complex structures assembled from DNA `bricks' \cite{Ke2012}, courtesy of P. Yin. (c) Tracks for a DNA-based motor on origami. Reprinted by permission from Macmillan Publishers Ltd: Nature Nanotechnology \cite{Wickham2012}, copyright 2012. (d) Basic displacement. (d.i) The initial state with incumbent (red) and substrate (blue) bound, and the invader (green) free in solution. (d.ii) The invader binds to the available toehold of the substrate. (d.iii) The invader competes with the incumbent for base pairs. (d.iv) The final state, in which the invader has completely displaced the incumbent.
\label{nanotech_examples}}
\end{figure*}
DNA is not restricted to static nanostructures, but can be used to create dynamically active nanoscale objects. Toehold-mediated strand displacement (TMSD), illustrated in Fig.\,\ref{nanotech_examples}\,(d), is used along with hybridisation to drive conformational changes in many such devices. TMSD involves the replacement of a strand (the {\it incumbent}) from a duplex with the {\it substrate}  by an alternative strand (the {\it invader}) that is complementary to an additional {\it toehold} of the substrate. Initially, the invader can bind to the exposed toehold, as shown in Fig.\,\ref{nanotech_examples}\,(d.ii) \cite{Yurke2003, Zhang_disp_2009, Srinivas2013}. The invader and incumbent can then compete for base-pairing, as illustrated in Fig.\,\ref{nanotech_examples}\,(d.iii). If, eventually, the incumbent loses all of its base pairs, it will detach and displacement is complete (Fig.\,\ref{nanotech_examples}\,(d.iv)). As well as being essential in dynamic nanotechnology, displacement may occur during the assembly of complex structures that initially form unintended bonds between the wrong strands.

The simplest devices are switches that respond to a change in experimental conditions. An iconic example are the `tweezers' of Yurke {\it et al.} \cite{Yurke2000}, which can be closed and opened by sequential addition  of `fuel' strands (which bind to and closes the tweezers)  and `antifuel' strands (which displace the tweezers from the fuel, opening them). Switches can allow large structures to open, close or undergo topological rearrangement \cite{Yan2002, Goodman2008, Andersen09, Han2010}.
 
DNA `walkers'  couple the mechanical change generated by DNA reactions to motion along extended tracks, analogous to biological motors such as kinesin and myosin. The earliest designs used sequential strand addition to generate unidirectional motion along  a series of binding sites (`stators') \cite{Sherman2004, Shin2004}. 
Solution conditions can also be modulated in different ways, for example through periodic exposure to light of different frequencies \cite{Liu2014}. Autonomous motors, which don't rely on external control, must catalyse the release of free energy \cite{Bath2007}. To meet this requirement, designs often couple motion to the hydrolysis of a nucleic acid strand that is not part of the walker \teo{itself. Hydrolysis can be achieved either with \cite{Bath2005, Bath2009,Wickham2011} or without \cite{Tian2005,Lund2010,Cha2014} the involvement of an additional enzyme. An} alternative  is to catalyse DNA hybridisation itself \cite{Green2008,Omabegho2009} if the reactants are present as metastable self-bonded hairpins. Many walkers destroy the track behind them \teo{in} the so-called `burnt bridges' approach. The Turberfield group have also demonstrated the possibility of autonomous motion on a track that can be reused \cite{Green2008,Bath2009}. In the majority of cases, hybridisation and displacement are central to walker operation, sometimes in combination with additional enzymatic reactions. An alternative is to use four-way branch migration rather than displacement \cite{Venkataraman2007, Muscat2011}.

A third branch of active DNA nanotechnology is computation. In 1994 Adleman showed
that  a Hamiltonian path problem could be solved using DNA  \cite{Adleman1994}. Subsequently, researchers have developed motifs for logical operation based on TMSD \cite{Seelig2006}, with the Winfree group having incorporated multiple gates into a decision-making `brain' \cite{Qian2011}, and Chen {\it et al.} having developed a general architecture for control networks \cite{Chen2013}.

\subsection{Applications of DNA nanotechnology}
Many nanotechnological systems are elegant proofs of principle, rather than being directly useful. The potential of DNA nanotechnology, however, is obvious. Seeman's original motivation was to use DNA crystals to conjugate proteins and facilitate crystallography; origami bundles and two-dimensional lattices have indeed been used to aid protein structure determination \cite{Berardi2011,Selmi2011}. Similarly, finite-sized nanostructures contain unique strands at well-defined locations. This allows  nanostructures to function as nanoscale breadboards for biophysical systems -- they have been used as backbones for plasmonic devices \cite{Kuzyk2012} and light-harvesting complexes \cite{Dutta2011}, to control \cite{Tseng2013,Minghui2013} and facilitate the study of \cite{Wilner2009,Fu2012,Fu2014} enzymatic reactions,  and to provide tracks for DNA motors \cite{Wickham2012, Tomov2013} (as shown in Fig.\,\ref{nanotech_examples}\,(c)). DNA nanostructures have also been designed to act as microscopic gene-detecting arrays \cite{Ke2008,Subramanian2011}. The freedom to design complex structures allows DNA to be used for other experimental components, including `handles', `frames' and `rulers' for single-molecule manipulation \cite{Endo2012,Pfitzner2013}, and harnesses for arrays of motor proteins \cite{Derr2012}. 

The therapeutic potential of DNA nanotechnology has long been recognised. Nanostructures can encapsulate molecules either covalently or non-covalently \cite{Douglas2012,Crawford2013}, with the aim of selectively releasing them at the surface of or inside \teo{specific} cells. Alternatively, nanostructures can coordinate biomolecules to mimic pathogens, potentially triggering stronger immune responses \cite{Li2011,Schuller2012,Liu2012} and allowing the design of novel vaccines \cite{Liu2012}. DNA computation may also prove to be most valuable in \teo{medical applications}, when its enormous parallel capacity and direct interaction with biomolecules will prove most advantageous. Towards this end, both the uptake of small nanostructures into cells \cite{Walsh2011,Li2011, Lee2012} and specific targeting of cancer cells by DNA-transported drugs \cite{Douglas2012} have been demonstrated. Large DNA nanostructures have been reported to be survive intact in cell lysate \cite{Mei2011},  and DNA-based devices and structures have been shown to be stable within {\it C. elegans} for hours or days, depending on the design \cite{Surana2013}. The Shih lab has also shown that lipid encapsulation of nanostructures can help to shield them from digestion by nucleases when this is an issue \cite{Perrault2014} and coating origami with virus capsid proteins has been shown to improve transfection into cells \cite{Mikkila2014}. Interacting ``robots" have even been demonstrated to perform computations based on TMSD  within a living organism \cite{Amir2014}.

Artificial walking devices could function as agents in molecular assembly lines,  incorporating some degree of decision-making. Preliminary work has demonstrated that DNA hybridisation can accelerate and template reactions \cite{He2010,McKee2012}, that walkers can make decisions at junctions \cite{Muscat2011,Wickham2012} and that walkers can selectively pick up gold nanoparticle cargo \cite{Gu2010}. 

\subsection{Understanding and optimising DNA nanotechnology}
If DNA nanotechnology is to be widely used, assembly processes and operation cycles must be understood and optimised. Simple self-assembling structures such as DNA tetrahedra form reliably when a solution  of reactants is rapidly cooled  \cite{Goodman2005}. Larger systems tend to be more complex, although recent work has shown that \teo{substantial} structures can form surprisingly well if the temperature is carefully chosen \cite{Sobczak2012,Myhrvold2013}. It is not obvious, however, why assembly can be so successful given previous failures with large colloidal structures \cite{Reinhardt2014}. It is also not clear why origami assembly can occur over a very narrow temperature window \cite{Sobczak2012}, with significant hysteresis \cite{Sobczak2012, Arbona2013}. Improving yields of larger structures by choosing sequences to minimise unintended cross-interactions has shown surprisingly little promise thus far \cite{Wei2012,Ke2012}, but origami assembly is very sensitive to the staple layout \cite{Ke2012b, Martin2012,Myhrvold2013}.
Even if an object appears to form well, some strands might be absent, possibly compromising its usefulness and mechanical properties \cite{Chen2014}.   Optimisation is even more important for dynamic nanotechnology, where relatively slow unintended reactions can compromise device operation \cite{Tomov2013}. For example, even if a walker has only a 5\% chance of detaching from the track at each step, most will fail to take twenty steps. As it stands, motors are slow compared to natural analogs,  and systematic approaches for improving their effectiveness or decision-making abilities are \teo{currently} limited.

Recent experiments have probed specific systems relevant to nanotechnology in detail \cite{Zhang_disp_2009,Sobczak2012,Myhrvold2013,Ke2012b,Martin2012,Tomov2013,Johnson-Buck2013,Tsukanov2013,Teichmann2014}. To interpret these results, generalise the resultant ideas and develop new principles, these systems must be modelled theoretically. To date, the main theoretical tool has been the nearest-neighbour model of DNA thermodynamics \cite{SantaLucia2004},  implemented in online tools such as NuPack \cite{Dirks2007}. The nearest-neighbour model functions at the level of  secondary structure ({\it i.e.,} lists of the base-pairing that is present in the system). The free energy of a given secondary structure can be estimated by summing contributions from each neighbouring `stack' of base pairs, plus contributions from end effects and enclosed loops. Whilst the nearest-neighbour model is extremely useful, it has several limitations. Firstly, it is a thermodynamic model with discrete states and hence has  no natural kinetics \cite{Srinivas2013}. Secondly, it does not explicitly represent  DNA structure, and struggles to describe complex interconnections, loops and `pseudoknots' \cite{Dirks2007} that often arise in nanotechnological systems. Finally, DNA mechanics is ignored, meaning that the effects of  forces and torques cannot be directly understood.   

Next, I will introduce coarse-grained models of DNA, mesoscale representations that can overcome the limitations of the nearest-neighbor description.  I will focus on oxDNA, a model explicitly designed for DNA nanotechnology. I will outline the model, before discussing previous applications in understanding TMSD in a variety of contexts. Finally, new results will be presented on displacement involving a two-footed DNA walker \cite{Bath2009}, highlighting the possibility of accelerating displacement through mechanical strain.

\section{Coarse-grained modelling of DNA nanotechnology}
Coarse-grained models provide a level of resolution between fully atomistic treatments  and secondary-structure descriptions like the nearest-neighbour model.
\teo{Atomistic treatments are generally too computationally expensive for nanotechnological applications, although some structural studies have been performed \cite{Oteri2011,Yoo2013}}. \teo{Coarse-grained} models represent individual nucleotides using a small number of interaction sites, which interact through effective potentials. If well parameterised, they can capture the known thermodynamic, structural and mechanical properties of DNA in a simple and naturally dynamical representation. 

The choice of interactions determines the accuracy and applicability of the model. A `bottom-up' approach is to fit interactions to reproduce correlation functions from more detailed atomistic simulations \cite{ Becker2007, Becker2009, Lankas2009, Sayar2009, Savelyev2010, Dans2010, Morriss-Andrews2010, Kikot2011, Savin2011, Gonzalez2013, He2013, Kovaleva2014, Maffeo2014, Korolev2014, Naome2014}, or data from experimentally determined structures \cite{Olson1998, Becker2009, Trovato2008, He2013} This procedure can be very effective, but it has limitations. Firstly, the resultant model is very dependent on the source data, which are usually primarily drawn from  duplex DNA, whereas nanotechnological systems involve ssDNA and hybridisation transitions (one exception has focused specifically on ssDNA only \cite{Maffeo2014}). Even if a wider variety of systems were to be used for parameterisation, it is not clear whether current atomistic representations  accurately describe isolated ssDNA and the hybridisation transition. Secondly, one cannot retain all features of a system when coarse-graining; \teo{`representability problems' arise \cite{Johnson2007}}. For a given set of coarse-grained degrees of freedom, the optimal potential for correlation functions will likely be distinct from that for thermodynamics. Related issues are relevant to potentials derived from {\it ab initio} calculations \cite{Hsu2012}.

\begin{figure}
\centering
\includegraphics[width=0.45\textwidth]{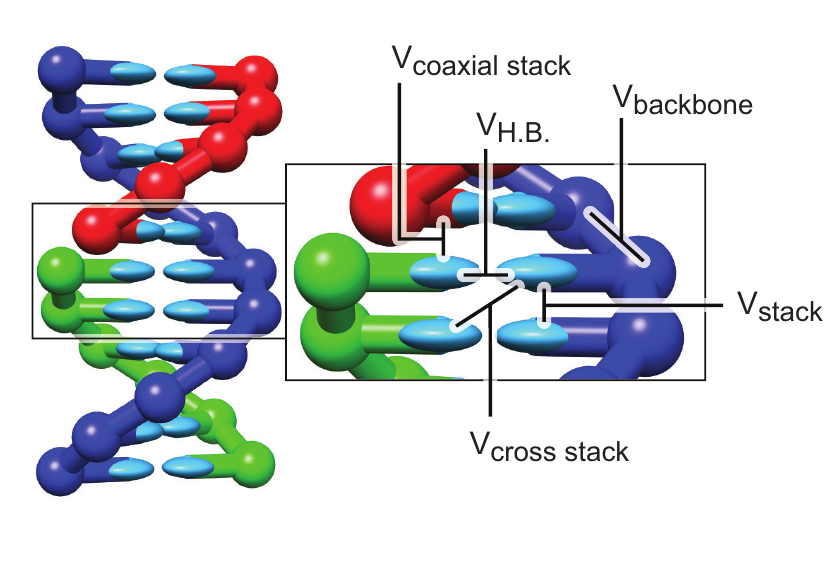}
\centering \caption{\footnotesize Illustration of the duplex structure and stabilising interactions with the oxDNA model. Spheres represent the backbone sites, and ellipsoids represent bases. Strands are coloured according to their identity. Coaxial stacking acts like a nearest-neighbour stacking interaction between bases that are not consecutive on the same strand. Excluded volume interactions (not illustrated here) also exist between all backbone and base sites. \teo{Reprinted with permission from T.E. Ouldridge, {\it et al.,} Optimizing DNA nanotechnology through coarse-grained modelling: A two-footed DNA walker, ACS
Nano 7, 2479. Copyright
(2013) American Chemical Society.}
}
\label{fig_model_DNA}
\end{figure}

Representability problems highlight the fact that coarse-grained models cannot be perfect, and hence should be designed with their purpose in mind. For nanotechnological systems, models must capture ssDNA, dsDNA and their interconversion. To date, this has been most successfully done with `top-down' approaches \cite{Drukker2001,Sales-Pardo2005,Starr2006,Tito2010,Knotts2007,Sambriski2009_BiophysJ,Hinckley2013, Svaneborg2012,Kenward2009,Linak2012,Araque2011,Niewieczerzal2009,Arbona2012b, Edens2012, Dasanna2013, Cragnolini2013,Doi2010, Mielke2005,Knorowski2011}. Here, physically-motivated interactions such as hydrogen-bonding and stacking are included and parameterised to reproduce overall thermodynamic, mechanical and structural properties of DNA. In this article I focus on oxDNA \cite{Ouldridge2011, Ouldridge_thesis, Sulc2012},  an attempt to incorporate thermodynamics as parameterised by the nearest-neighbour model into a description with continuous degrees of freedom that captures structural and mechanical properties of DNA. Due to the emphasis placed on capturing the duplex formation transition in the development of oxDNA, it has been used to study nanotechnological systems more extensively than other models. However,  duplex and hairpin formation transitions have been systematically investigated by a number of groups \cite{Sales-Pardo2005, Starr2006, Tito2010, Allen2011, Sambriski2009_BiophysJ, Prytkova2010,Hoefert2011, Schmitt2013, Hinckley2013,Linak2012, Edens2012, Cragnolini2013, Hinckley2014}, nanostructure conformation has been studied by Bombelli {\it et al.} and Arbona {\it et al.} \cite{Bombelli2008,Arbona2012b}, and nanostructure assembly  \cite{Arbona2012, Svaneborg2012b, Ouldridge2009}  and displacement  \cite{Svaneborg2012c} have been demonstrated with much simple models of DNA. Large-scale asssembly of DNA-coated objects has also been achieved with some simpler coarse-grained models \cite{Starr2006,Knorowski2011,Li2012}. It is worth noting that similar top-down \cite{Hyeon2005, Cao2006, Jost2010, Ding2008, Pasquali2010, Dickson2011, Cragnolini2013, Denesyuk2013} and bottom-up \cite{Jonikas2009,Das2009,Paliy2010,Xia2013} models exist for RNA.

\subsubsection{oxDNA}

oxDNA  treats each nucleotide as a 3-dimensional rigid body. 
The potential energy of a configuration is given by 
\begin{eqnarray}
  V =  \sum_{\left\langle ij \right\rangle} \left( V_{\rm{b.b.}} + V_{\rm{stack}} +
V^{'}_{\rm{exc}} \right) + \nonumber \\
      \sum_{i,j \notin {\left\langle ij \right\rangle}} \left( V_{\rm HB} +  V_{\rm{cr.st.}}  +
V_{\rm{exc}}  + V_{\rm{cx.st.}} \right).
 \label{eq_hamiltonian}
\end{eqnarray}
Here the  first sum runs over all consecutive pairs of nucleotides on a strand, 
and the second sum over all remaining pairs. The interactions represent hydrogen bonding ($V_{\rm HB}$), cross stacking ($V_{\rm{cr.st.}}$), coaxial stacking ($V_{\rm{cx.st.}}$),
nearest-neighbour stacking  ($V_{\rm{stack}}$), excluded volume ($V_{\rm{exc}}$ or $V^{'}_{\rm{exc}}$) and backbone chain connectivity ($V_{\rm{b.b.}} $). These interactions are shown schematically in Fig.\,\ref{fig_model_DNA}, and are discussed in detail elsewhere \cite{Ouldridge2011,Ouldridge_thesis,Sulc2012}. Importantly, attractive interactions depend explicitly on the relative orientations of nucleotides, allowing the anisotropic nature of bases to play a role.

Hydrogen-bonding and stacking interactions drive the formation of duplexes with helical structure from single strands that are relatively more disordered. oxDNA reproduces the thermodynamic, mechanical and structural changes associated with this transition, under high salt conditions. In particular, oxDNA provides a good representation of duplex melting temperatures, melting transitions widths, self-complementary hairpin stability, duplex elastic moduli  and the short persistence length of single strands (details are provided in Refs. \cite{Ouldridge2011,Ouldridge_thesis,Sulc2012}). Our group uses the ``Virtual Move Monte Carlo" (VMMC) algorithm (the variant in the appendix of Ref. \cite{Whitelam2009}) to calculate model thermodynamics. Dynamical properties require an additional choice of model kinetics; we use Langevin \cite{Davidchack2009} and Andersen-like \cite{Russo2009} thermostats. To improve sampling, our group has made extensive use of umbrella sampling (US) \cite{Torrie1977} \teo{for thermodynamic averages and forward flux sampling (FFS) \cite{Allen2005,Allen2009} for kinetic studies}. 

Several important simplifications are inherent in the model. Firstly, oxDNA  was fitted at a salt concentration of  [Na$^+]=0.5$\,M where electrostatics is strongly screened -- the repulsion of phosphates is therefore incorporated into the backbone excluded volume for simplicity. A recent study by the Pettitt group \cite{Wang2014} has explored the possibility of incorporating a Debye-H\"{u}ckel description of electrostatics, allowing lower salt conditions to be simulated, but the results reported in this work do not include this term. Secondly, model duplexes are symmetric, meaning that both grooves of the helix are the same size. \teo{Such a simplification may be of relevance when systems are extremely sensitive to geometric details; for example, the stress imposed on helices by crossovers at origami junctions will be affected by grooving.} \teo{Thirdly, as nanotechnology typically involves low concentrations of DNA, oxDNA takes the partial pressure of strands to be zero. Therefore simulations of DNA with implicit solvent in the canonical ensemble are appropriate for comparisons to typical experimental systems at constant temperature and pressure \cite{Ouldridge_bulk_2012}. Consistent with this picture, we interpret free energies measured in simulations as Gibbs (rather than Helmholtz) free energies.}  Finally, the Langevin and Andersen-like thermostats do not incorporate collective hydrodynamic motion, and low friction coefficients are typically used to accelerate dynamics. Given these and other simplifications, it is important to identify the underlying cause of any phenomena observed in simulation, to ensure that they arise from real DNA physics rather than artefacts of the model or dynamical algorithm. 

\section{Insights into strand displacement from oxDNA}
 I now outline how oxDNA has been used to understand basic TMSD, and then to explore variants relevant to nanotechnology.
 \begin{figure}
\centering
\includegraphics[width=0.32\textwidth,angle =-90]{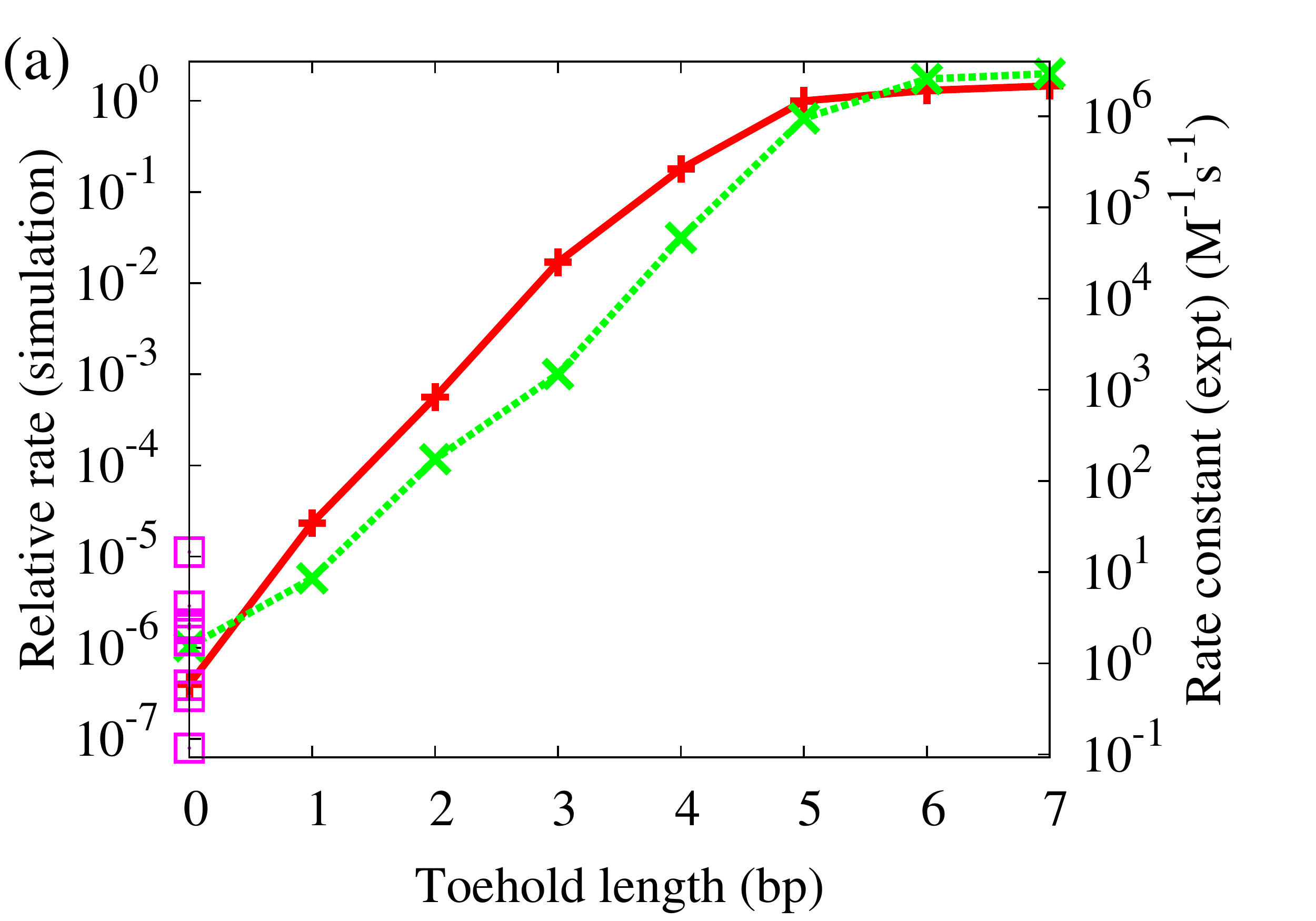}
\includegraphics[width=0.32\textwidth,angle =-90]{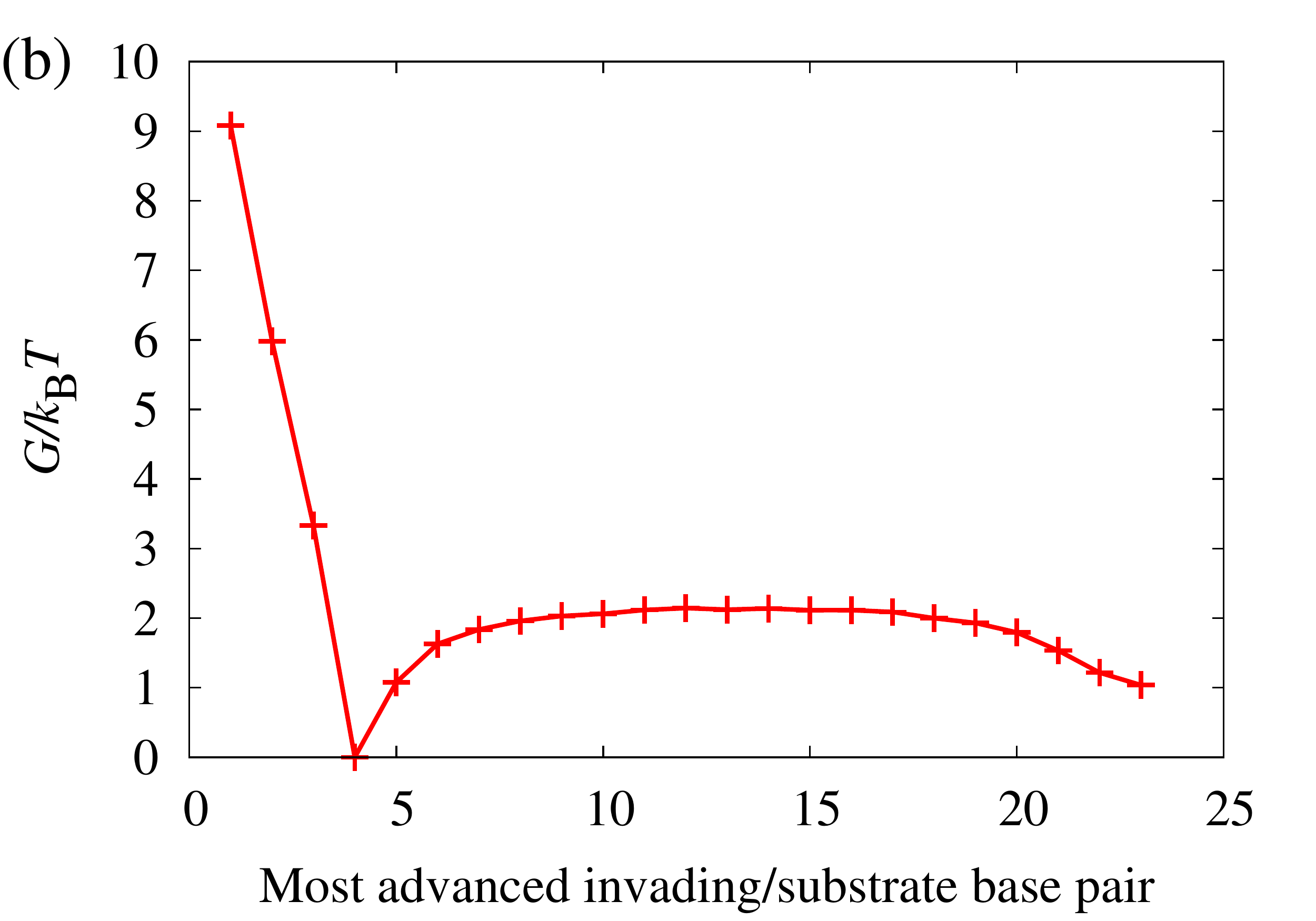}
\centering \caption{\footnotesize (a) Experimentally measured rate constants as a function of length for an average-strength toehold \cite{Zhang_disp_2009} (green `x' symbols), compared to the relative displacement rates found for oxDNA (red `+' symbols, \teo{reported relative to the 5 base-pair case}) \cite{Srinivas2013}. Purple squares show the new estimates of zero-toehold displacement rate for oxDNA presented in this work (detailed in the Supplementary Material), showing substantial measurement uncertainty. 
\teo{Although not quantified here, this uncertainty is much smaller for longer toeholds, when the process is easier to sample \cite{Srinivas2013}. The minor variation between rates for long toeholds demonstrates the small error in this limit}. (b) Free energy ($G$) profile of displacement in oxDNA as a function of the most advanced invader/substrate base pair for a four-base toehold. \teo{Base pair `1' is at the far end of the toehold from the displacement domain, and base pair 4 is immediately adjacent to the displacement domain} \cite{Srinivas2013}. Typical errors on these measurements are smaller than the `+' symbols, \teo{as estimated through comparison of independent simulations \cite{Srinivas2013}}. 
\label{fig_basic_disp2}}
\end{figure}
\begin{figure*}
\centering
\includegraphics[width=0.9\textwidth]{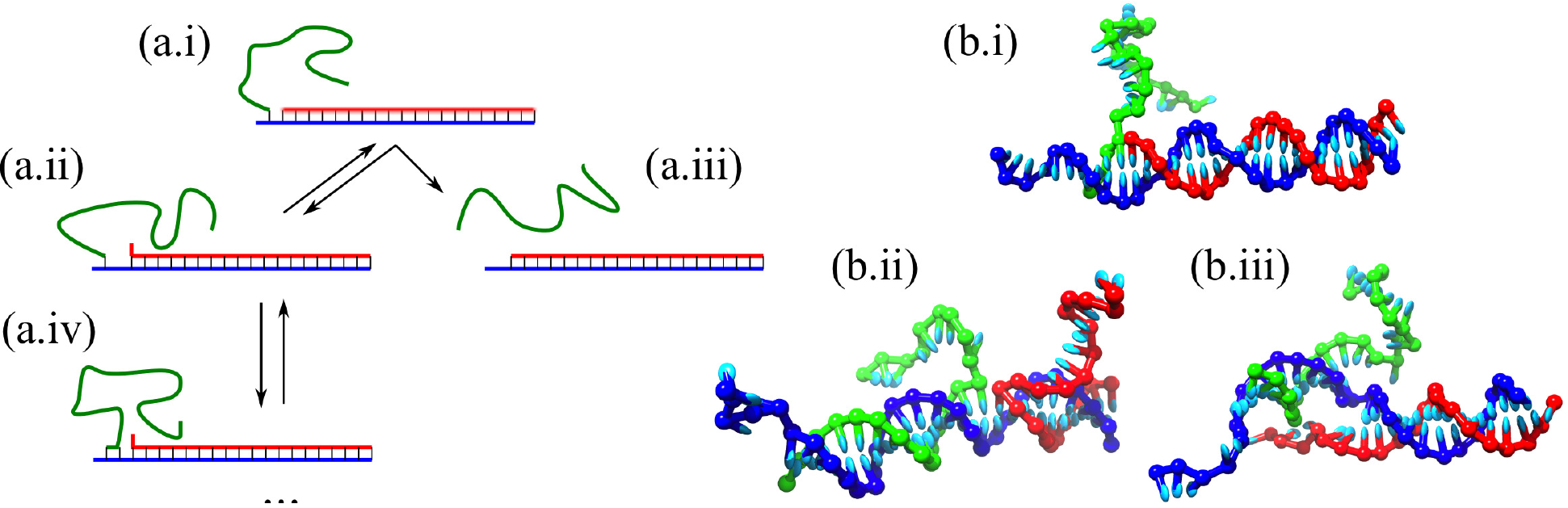}
\centering \caption{\footnotesize Representations of displacement at two levels of detail. In both cases the substrate is blue, the invader green and the incumbent red. (a) Illustration of a simple step-by-step model of displacement for a one-base toehold. Starting from the toehold-bound state (a.i), either a base pair between incumbent and substrate can break (a.ii) or the toehold base pair can break, leading to detachment (a.iii). From state (a.ii), either the invader can bind to the newly revealed substrate base (a.iii), or the incumbent can rebind, returning us to the initial state (a.i). This process can be repeated for every base in the displacement domain. (b) Displacement in oxDNA (for a three-base toehold). (b.i) The toehold-bound state. (b.ii) A later stage of displacement. (b.iii) A later stage of displacement in which the invader and incumbent duplexes have unstacked at the displacement junction.
}
\label{fig_basic_disp}
\end{figure*}

\subsection{Basic displacement}
\label{basic disp section}

TMSD was experimentally characterised by Zhang and Winfree \cite{Zhang_disp_2009}. These authors demonstrated that for short toeholds, displacement rate increases exponentially with toehold length, before plateauing in the long toehold limit. The results, a subset of which are reproduced in Fig.\,\ref{fig_basic_disp2}\,(a), show that displacement accelerates by a factor of  $\sim 10^{6.5}$ as the toehold is increased from 0 to 15 base pairs (for a displacement domain of 20 base pairs, at 25$^\circ$C and with a high salt concentration of [Mg$^{2+}$]$ \approx 10$\,mM). The overall shape of the graph is unsurprising. At the low strand concentrations used in Ref. \cite{Zhang_disp_2009}, the three-stranded intermediate (including states depicted in Figs.\,\ref{fig_basic_disp}\,(a.ii), (a.iv) and (b)) is  short-lived and the reaction is effectively  second order. In this limit, the reaction rate constant can be modelled as
\begin{equation}
k_{\rm eff} = k_{\rm bind} P_{\rm suc},
\end{equation}
where $k_{\rm bind}$ is a toehold binding rate constant and $P_{\rm suc}$ is the probability that displacement succeeds (as opposed to the invader detaching) once the toehold is formed. For very short toeholds it is reasonable that \teo{$P_{\rm suc} \ll 1$}. Increasing the toehold length increases the toehold stability, and hence  $P_{\rm suc}$, exponentially. Eventually this increase with toehold length saturates when $P_{\rm suc}\sim 1$. Given that $k_{\rm bind}$  would be expected to be relatively weakly length-dependent, $k_{\rm eff}$ therefore plateaus at this point.

A simple argument, however, would suggest that a single-base toehold should have a success rate greater than 1\%, thereby limiting the possibilities for increasing $k_{\rm eff}$ by increasing  the toehold length and hence $P_{\rm suc}$. Consider the system illustrated in Fig.\,\ref{fig_basic_disp}\,(a), with a toehold of one base, and assume that displacement is a random walk in which base pairs at the junction can break and then be replaced by base pairs with the competing strand, as shown in Fig.\,\ref{fig_basic_disp}\,(a).  From the toehold-bound state, two things can happen -- either the toehold detaches, or the first base pair in the incumbent/substrate breaks. If the latter, the invader then takes a base pair from the incumbent 50\% of the time, \teo{otherwise the system returns to the initial state}. If toehold detachment (involving the disruption of a single base pair), and breaking of base pairs at the junction have similar rates, the probability that the invader manages to take the first step is then approximately 1/3. From here, 19 more displacement  steps are required, whereas the initial state is one backwards step away. From the statistics of random walks, we obtain $P_{\rm suc} \sim \frac{1}{3}\frac{1}{20} \sim 0.02$ for single-base toeholds. Given this estimated value of $P_{\rm suc} $ for a single-base toehold, it is difficult to justify the extent of the experimentally observed slowdown for shorter toeholds. 

The displacement process has been simulated \teo{with} oxDNA, as outlined in Ref. \cite{Srinivas2013}. The results shown in \teo{Fig.\,\ref{fig_basic_disp2}\,(a)} agree  well with those obtained by Zhang and Winfree. The key point, however, is not the quality of agreement, but that oxDNA also violates the simple argument outlined above which suggests only a modest slowdown for short toeholds. By studying oxDNA, we can then hope to explain the experimental results. Explanations for this behaviour, violations of two assumptions of our simple argument, are given discussed in detail Ref.\,\cite{Srinivas2013}. 

Firstly, junction migration is not an unbiased random walk. As junction migration is initiated, a second single-stranded section is generated at the displacement junction, as can be seen in \teo{Fig.\,\ref{fig_basic_disp}\,(b)}. This second overhang is thermodynamically unfavourable, due to steric exclusions at the junction, and consequently biases the system against initiating and proceeding with junction migration. It therefore reduces the probability of successful displacement given toehold binding. A free-energy profile
 indicating a thermodynamic cost for initiating displacement is shown in \teo{Fig.\,\ref{fig_basic_disp2}\,(b)} (the free energy as a function of an order parameter, $G(Q)$, is a measure of the probability that the system exists in state $Q$, $P(Q)$: $G(Q) = -k_{\rm B} T \ln P(Q) + {\rm const}$).
 
  Secondly, simulations show that junction migration is far more complex than the breaking of a single base pair in the toehold, necessarily involving the disruption of more stacking interactions and a greater structural rearrangement. Junction migration is then intrinsically slow compared to the disruption of toehold base pairs, again reducing the probability of displacement as opposed to detaching from the toehold.

The Winfree group have used the thermodynamic stability of `frozen' displacement intermediates to explore the possibility of the penalty associated with dangling ssDNA at the junction, confirming its existence and estimating a value of $\sim 3k_{\rm B}T$, slightly larger than found for oxDNA \cite{Srinivas2013}. Analysis of a simple secondary-structure based model \cite{Srinivas2013} suggests that such an impediment provides only a partial explanation of the \teo{degree to which longer toeholds can accelerate displacement}. \teo{As a consequence, the second factor identified by oxDNA as inhibiting displacement (slow junction migration) seems necessary to explain the wide range of rates as a function of toehold length.}

\subsection{Displacement involving mismatches}
\begin{figure}
\centering
\includegraphics[width=0.32\textwidth, angle=-90]{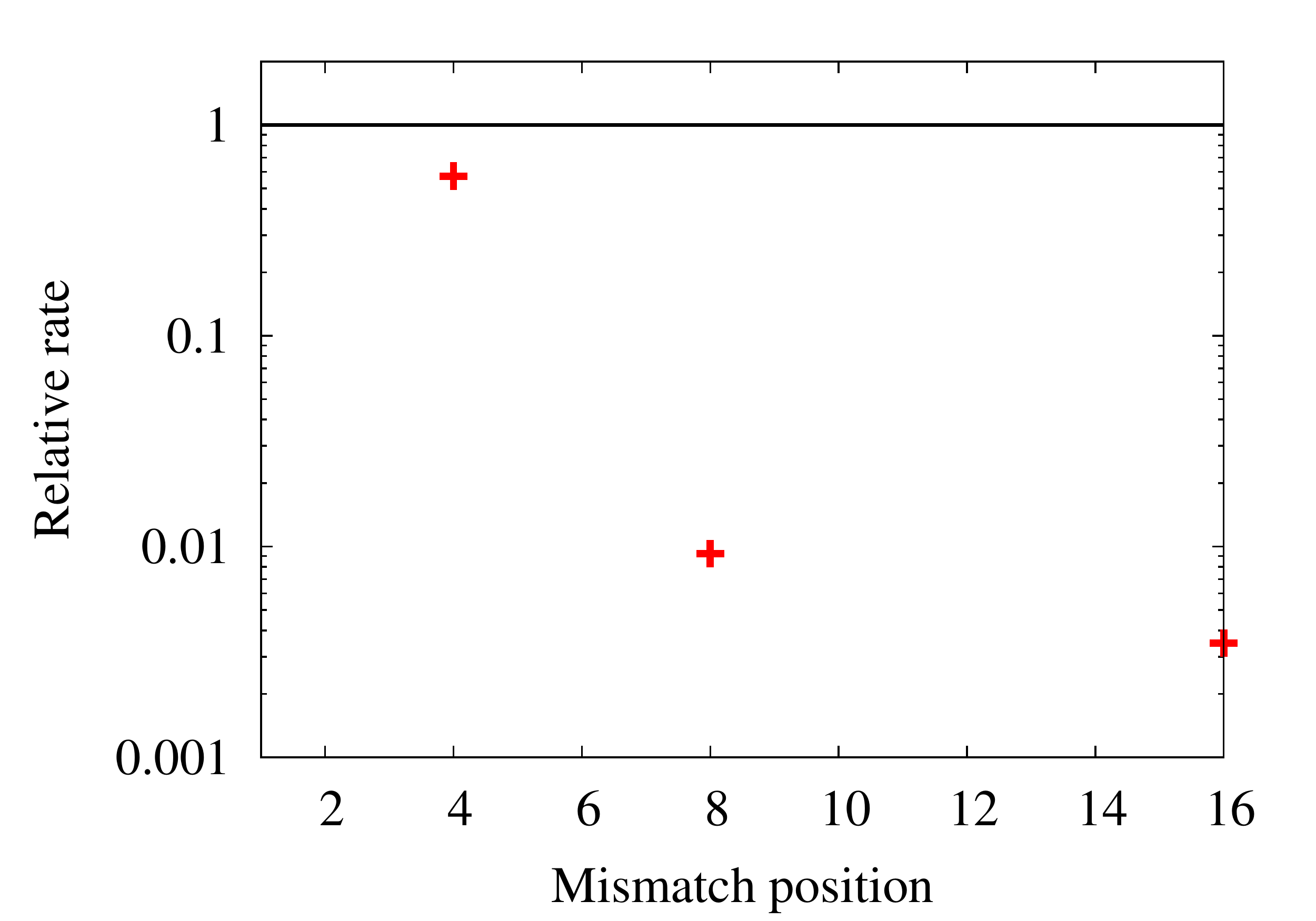}
\centering \caption{\footnotesize Relative rates of displacement as a function of the position of a mismatch created during displacement (as predicted by oxDNA) \cite{Machinek2014}. The position is defined with respect to the far end of the displacement domain (which has a length of 16 base pairs in this system) -- a mismatch at position `16' is immediately adjacent to the toehold, and a mismatch at position `4' is near the far end of the displacement domain. Values are reported relative to the rate for a perfectly matched invader, and the all points are estimated to have an error less than or equal to a factor of 2, \teo{as estimated through independent simulations} \cite{Machinek2014}.}
\label{fig_mm_disp}
\end{figure}

\begin{figure*}
\centering
\includegraphics[width=0.8\textwidth]{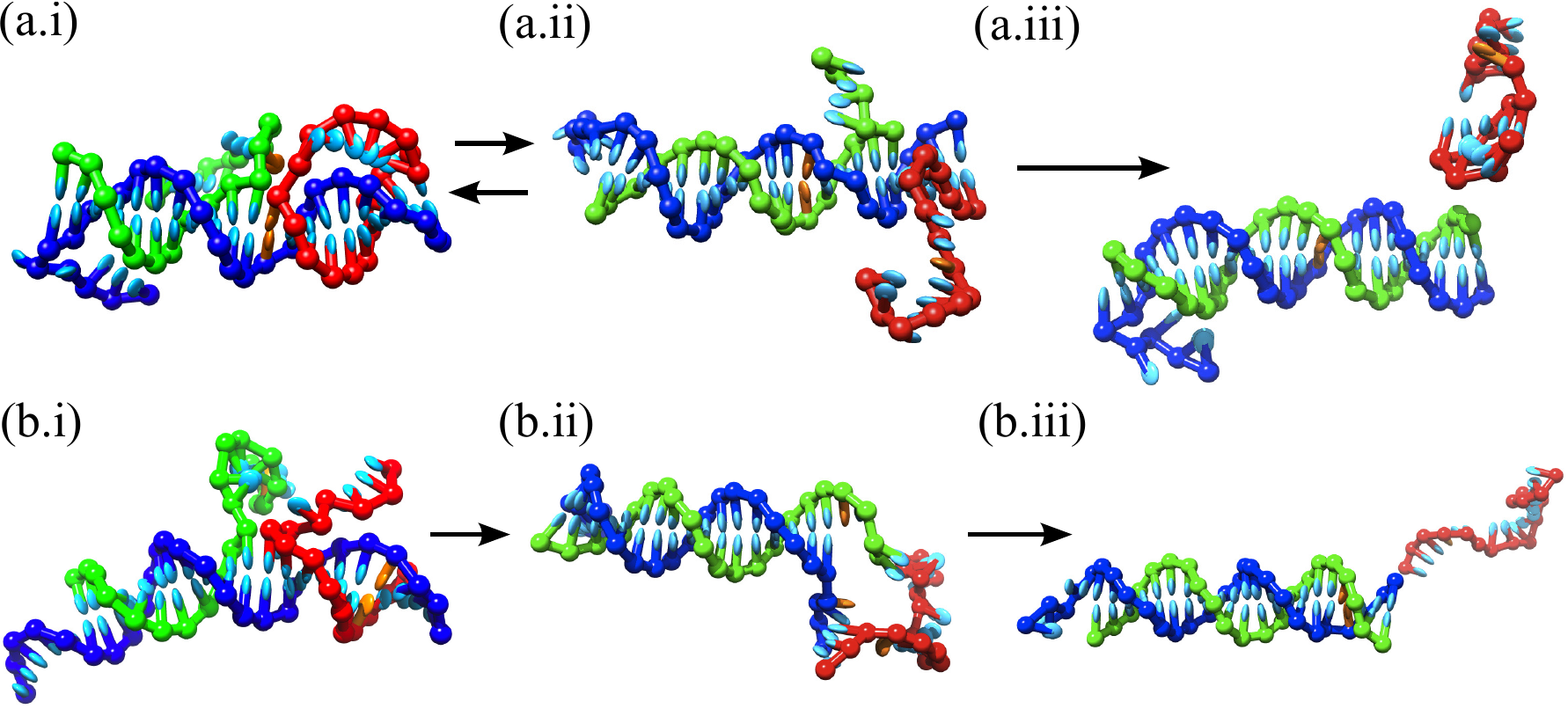}
\centering \caption{\footnotesize Distinct pathways for displacement involving the creation of a mismatched base pair. The incumbent is shown in red, the invader in green and the substrate in blue. The location of the mismatch is highlighted by the orange-coloured bases. (a) Creating a mismatch in the middle of the displacement domain. The typical pathway involves enclosure of the mismatch (a.ii) prior to detachment of the incumbent and the completion of displacement (a.iii). (b) Creating a mismatch four bases from the end of the displacement domain typically involves detachment of the incumbent (b.ii) prior to enclosure of the mismatch (b.iii). 
}
\label{fig_mm_disp2}
\end{figure*}
The plateau in displacement rate as toehold length increases limits the maximal selectivity for intended processes over leak reactions.
Modifying displacement to reduce its success probability would enlarge the regime in which rate grows with toehold stability, and hence the dynamic range and maximal selectivity. Authors have included physical separation of the toehold from the displacement domain \cite{Genot2011} or forced the invading strand to create unfavourable `mismatched' base pairs \cite{Zhangmm2012,Jiang2014} to achieve this. The principle is generally to slow displacement by decreasing the overall free-energy gain $|\Delta G^{\rm disp}|$ of the reaction.

OxDNA has been used to simulate displacement in which a C-G base pair is replaced by a C-C mismatch \teo{as displacement proceeds}
 \cite{Machinek2014}. Mismatches were considered at the start, in the middle, and four base pairs from the end of a 16-base displacement domain, for a toehold of 5 bases. In each case, the mismatch  destabilises the final invader/substrate duplex by approximately $\Delta\Delta G^{\rm disp}_{\rm mm}=5.9 k_{\rm B}T$. We might then na\"{i}vely expect a rate reduction by a factor
 \begin{equation}
\frac{k_{\rm eff}^{\rm mm}}{k_{\rm eff}^{\rm perfect}} \approx \exp(-\Delta\Delta G^{\rm disp}_{mm}/k_{\rm B}T) \approx \frac{1}{350}
\end{equation}
The results, taken from Ref.\,\cite{Machinek2014} and shown in Fig.\,\ref{fig_mm_disp}, are initially surprising. The early mismatch does indeed cause a slowdown by  $\sim 350$, and the middle mismatch by a factor $\sim 100$. The late mismatch, however, has almost no influence. This result must mean that $P_{\rm suc}$ is high despite the late mismatch. The fact that the mismatch is far from the toehold certainly helps -- after encountering the mismatch, the junction must migrate 12 steps backwards before the invader is bound only to the toehold, raising the probability that displacement will occur anyway despite the impediment. However, mismatches are so destabilising ($5.9 k_{\rm B}T$ in this case) that it is still difficult to see how enclosing it and subsequently completing junction migration would not be slow compared to returning to the toehold-only state. The resolution of this paradox is the existence of an alternative pathway, in which the incumbent strand spontaneously detaches from the substrate at a stage when the invader has not yet enclosed the mismatch. \teo{This pathway should be contrasted with the  standard picture of base-by-base displacement \cite{Srinivas2013}}. The invader/substrate base pairs beyond the mismatch are then formed at a later stage, when there is no competition from the incumbent, and the full penalty of $\Delta\Delta G^{\rm disp}_{\rm mm}$ is not manifest in the displacement reaction rate. This alternative pathway is illustrated schematically, and contrasted with a displacement pathway which involves enclosing the mismatch, in Fig.\,\ref{fig_mm_disp2}. 

Spontaneous detachment involves breaking a number of base pairs. When the mismatch is late in the displacement domain, it is feasible for the incumbent to detach spontaneously when the invader reaches the mismatch location. \teo{This process is somewhat analogous to the detachment of a short toehold, as discussed in Section \ref{basic disp section}. As highlighted in Section \ref{basic disp section}, toehold detachment is relatively fast compared to completing displacement; this explains why spontaneous detachment involving the disruption of several base pairs can be a kinetically favoured pathway, even when the most stable final state involves enclosure of the mismatch by the invader/substrate duplex.} 

As the mismatch is moved towards the start of the displacement domain, the number of base pairs that must break spontaneously grows and the rate of detaching in this way is exponentially suppressed. Eventually, spontaneous detachment is so slow that it is no longer faster than simply continuing displacement in the conventional base-by-base manner, and enclosing the mismatch. For oxDNA, both the middle and early mismatches are in this regime and hence they feel the effect of $\Delta\Delta G^{\rm disp}_{\rm mm}$. \teo{The difference between middle and early mismatches arises not from differences in the rate at which displacement occurs given toehold binding, but actually due to slower toehold detachment for the middle mismatch \cite{Machinek2014}}.

In general, oxDNA predicts that  mismatches formed at the start of a displacement domain will heavily suppress displacement rates. Rates should initially rise slightly as the mismatch is moved towards the far end of the domain (away from the toehold). At some point, the spontaneous melting pathway will become \teo{relevant and the overall rate will rise} rapidly, eventually plateauing at the perfectly-matched value. Recent experiments have demonstrated exactly this behaviour \cite{Machinek2014}, showing that  oxDNA is a powerful tool for probing DNA reaction dynamics.

These results emphasise that mismatches should be placed at the start of the displacement domain to slow displacement and thus suppress the rate of leak reactions. Recent work has considered using displacement to resolve single-nucleotide polymorphisms in genes \cite{ Li2002, Subramanian2011,Zhang_disp_2012, Chen_disp_2013}; oxDNA suggests that the simplest kinetic methods might struggle to detect mutations at the end of a displacement domain \cite{Li2002, Subramanian2011}, although modified approaches might fair better \cite{Zhang_disp_2012, Chen_disp_2013}. More generally, the results illustrate a mechanism for modulating reaction kinetics by orders of magnitude whilst maintaining approximately the same overall reaction thermodynamics. A corollary is that reverse reactions will be far faster for the late mismatch, which may also be important when designing nanotechnological systems.

\subsection{Displacement during the cycle of a burnt-bridges motor}
\begin{figure}
\centering
\includegraphics[width=0.45\textwidth]{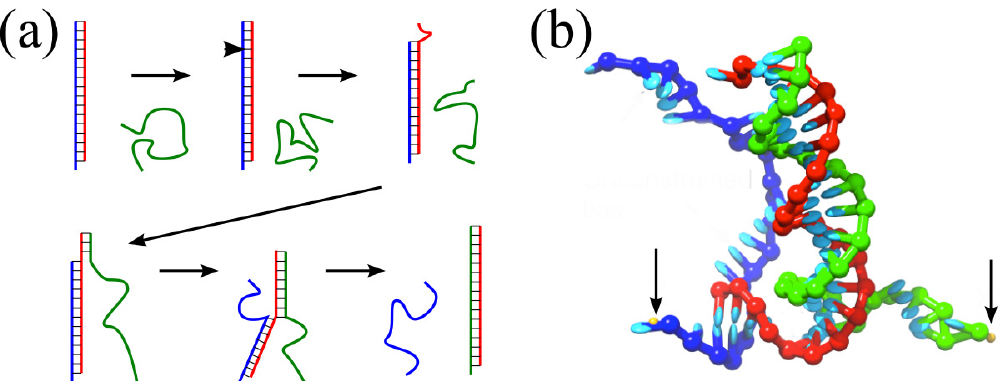}
\centering \caption{\footnotesize The burnt-bridges motor introduced in Ref. \cite{Bath2005}. (a) A schematic of of motor operation. The cargo (red) is initially bound to the first stator (blue). A nicking enzyme cuts this stator, revealing a toehold (six bases in this case) for an alternative stator (green) which can displace the original, resulting in a step of the cargo strand. (b) An intermediate stage of stepping as represented by oxDNA. The two stators are anchored at points indicated by the vertical arrows. From this figure it is clear that the final stages of displacement involve subjecting the motor to considerable tension as it stretches between the two anchoring points.
}
\label{fig_motor_bb}
\end{figure}

\begin{figure}
\centering
\includegraphics[width=0.45\textwidth]{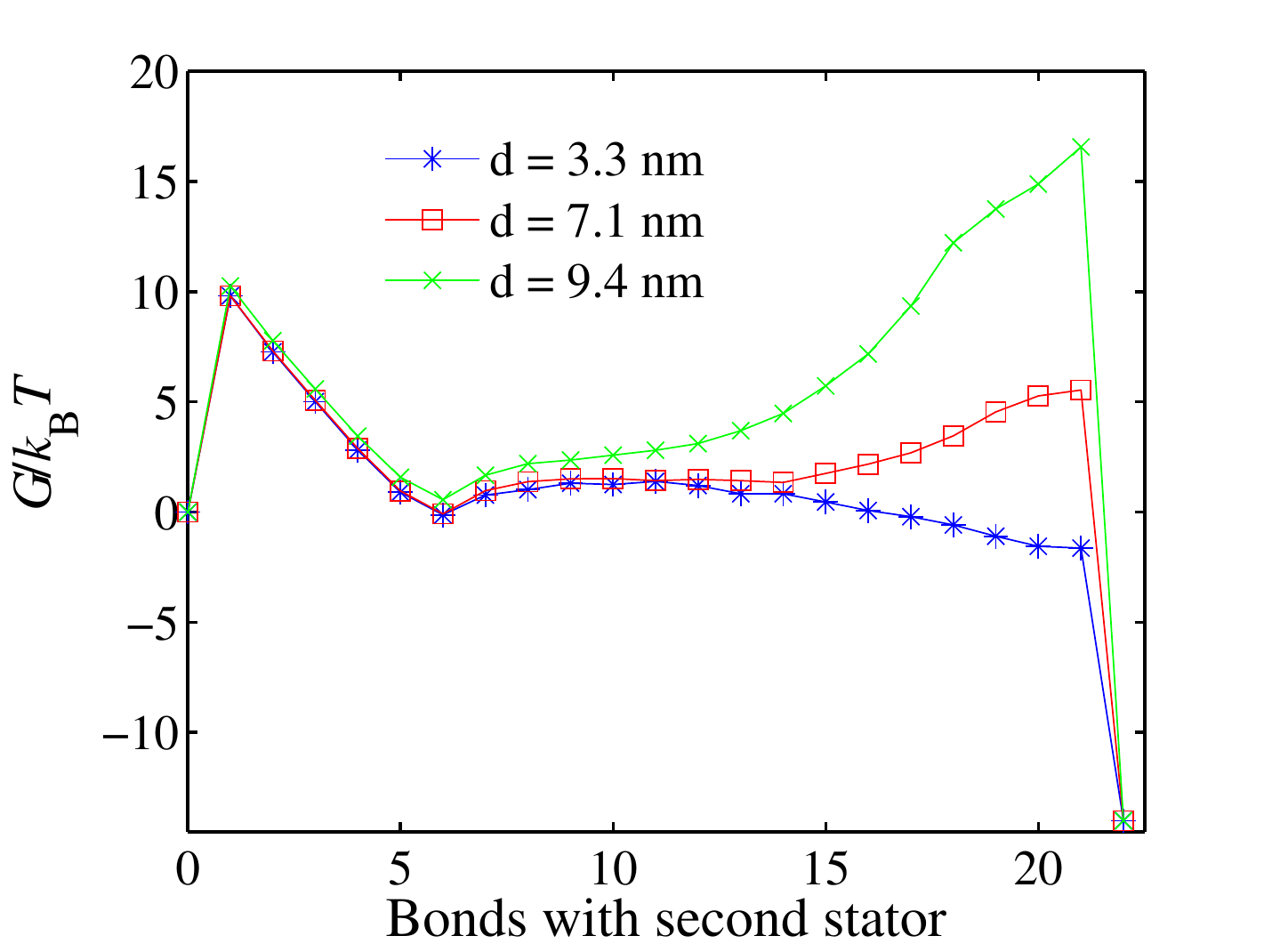}
\centering \caption{\footnotesize Free-energy ($G$) profiles  as a function of the number of bonds with the second stator during the displacement stage of the cycle of a burnt-bridges motor. Profiles are shown for three different separations between the anchoring points of the stators. \teo{Reprinted with permission from P. \v{S}ulc {\it et al.,} Simulating a burnt-bridges DNA motor with a coarse-grained DNA model, Natural Computing 1-13. Copyright (2013) Springer Science and Business Media \cite{Sulc_walker_2012}.}
}
\label{fig_motor_bb2}
\end{figure}
OxDNA has been used  \cite{Sulc_walker_2012} to model a burnt-bridges DNA motor designed by Turberfield and coworkers \cite{Bath2005,Wickham2011,Wickham2012}. The motor is illustrated in Fig.\,\ref{fig_motor_bb}\,(a). I will focus on the mechanism by which the motor strand (or `cargo') steps from one stator \teo{(a single-stranded binding site anchored to the track)} to the next via displacement - a simulation snapshot of this process is shown in Fig.\,\ref{fig_motor_bb}\,(b). The free-energy profiles of displacement, for various distances $d_s$ between the stators, are shown in Fig.\,\ref{fig_motor_bb2}. As $d_s$ is increased, toehold formation becomes slightly less favourable, due to a greater loss of entropy associated with toehold binding when the stators are further apart. Far more significant, however, is the rise in free energy at later stages  that can be seen for large $d_s$. As displacement progresses, fewer bases are available to stretch across  the gap between the points at which the stators are anchored, as shown in  Fig.\,\ref{fig_motor_bb}\,(b). Displacement must then work against the tension of these stretched bases, meaning that the free energy rises as displacement progresses.

Some caution must be exercised when using free-energy profiles to infer kinetics. \teo{If the low-dimensional order parameter used is not a perfect reaction coordinate, signatures of kinetic effects may be hidden. Indeed, the relative difficulty of junction migration highlighted in Section \ref{basic disp section} is not easily identified in free-energy profiles at the level of base pairs. }
Nonetheless,  large stator separation will clearly frustrate motor stepping, primarily due to lower rates of displacement once the toehold is formed, rather than slow 
toehold formation. Moderately large stator separations could then be used \teo{like early mismatches} to reduce displacement success probabilities, thereby increasing the toehold length required to saturate $P_{\rm suc}$ at unity and  suppressing leak reactions relative to the rate in this long-toehold limit. This \teo{approach} would be particularly useful at junctions where decisions must be made \cite{Wickham2012}.   If  $d_s$ primarily influenced binding rates, rather than success probabilities,  larger $d_s$ would not help to discriminate between toeholds but would slow all reactions equally.  

It is worth noting that slowing displacement by increasing $d_s$ alters reaction kinetics without changing the overall $\Delta G$ of reaction. \teo{This is similar to changing the location of a mismatch formed between invader and substrate, but unlike increasing the number of mismatches}. This is because the initial and final states, in which the cargo is bound to one stator only, are not $d_s$-dependent.  Large values of $d_s$ can in principle be used at an arbitrarily large number of stages for the same cargo molecule without destabilising the final product. This advantage of modulating kinetics with $d_s$ must  be weighed against the fact that the motif is limited to sequential reactions on a surface. 

\section{Recovery of a two-footed walker: an example of enhanced displacement}

\begin{figure*}
\centering
\includegraphics[width=0.9\textwidth]{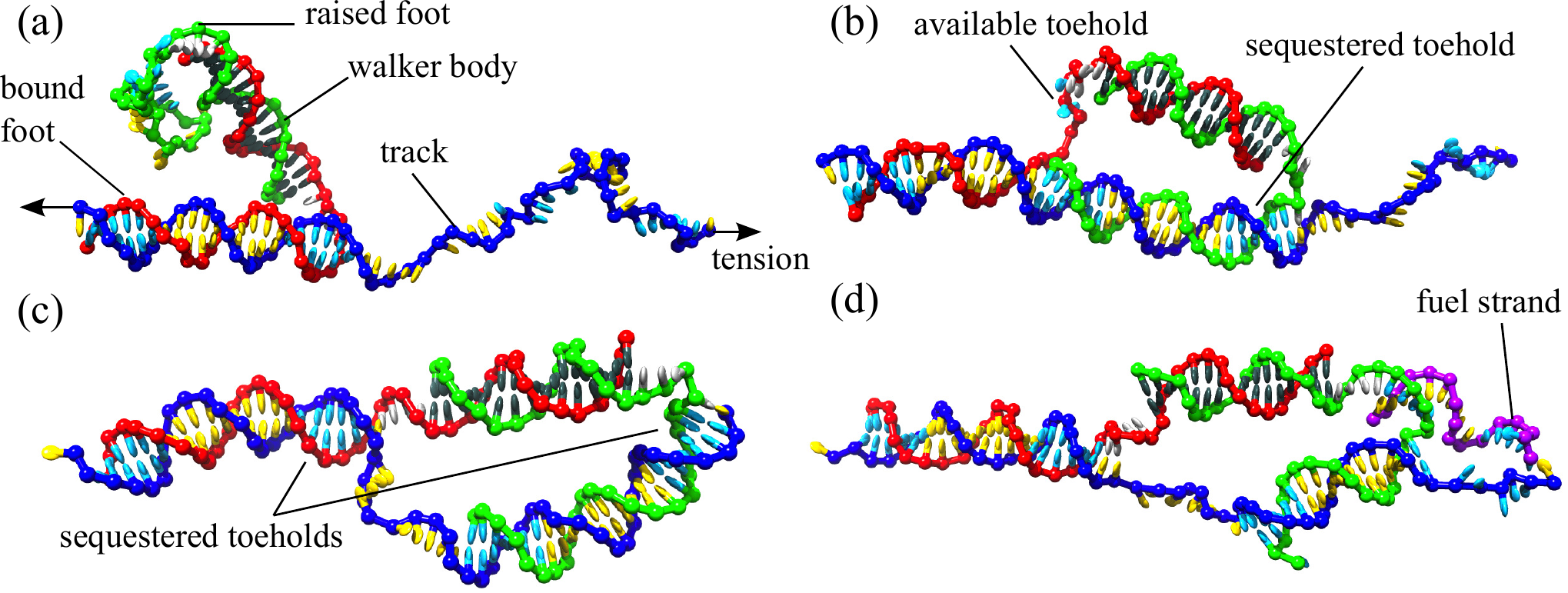}
\centering \caption{\footnotesize 
States relevant to the operation of a two-footed DNA walker. The backbone of the track is coloured blue, the two feet red and green and the fuel purple. Bases are coloured according to their `domain' -- sky blue bases are toehold sequences, and yellow bases constitute the longer binding domains between track and foot. Dark green bases form the duplex that holds the two feet together. The system has an inherent asymmetry which allows the definition of forward and backward directions \cite{Ouldridge_walker_2013}; in this figure, the forward direction is to the right. (a) Walker with one foot bound to the back site of a three-site track. (b) Walker bound to two adjacent sites on a three-site track, revealing a toehold for the fuel to raise the back foot. (c) Overstepped walker with sequestered toeholds. (d) Fuel displacing the track from the foot, prior to recovery of the walker from the overstepped state. In all figures, the track is subject to a tension of 14.6\,pN in the horizontal direction. 
\label{fig_walker}}
\end{figure*}

\begin{figure}
\centering
\includegraphics[width=0.32\textwidth, angle =-90]{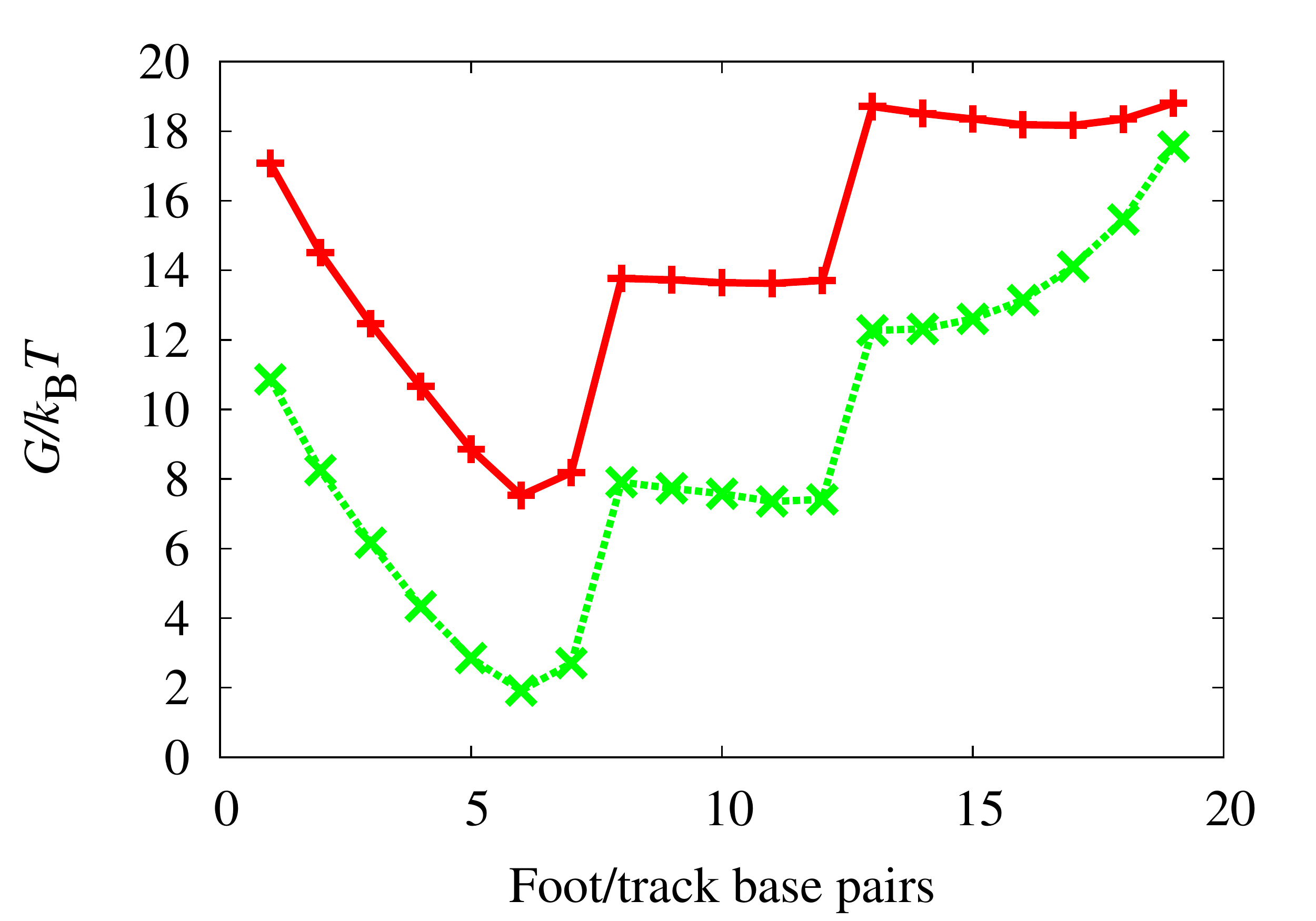}
\centering \caption{\footnotesize  Free-energy profile $G$ for the raising of the front foot of an overstepped walker, given at least one base pair between an invading fuel strand and the foot. Data is shown for systems both with (green `x' symbols) and without (red `+' symbols) a tension of 14.6\,pN applied the track. Free energies  are measured relative to a reference state in which the fuel is yet to bind. Increasing displacement corresponds to reduced numbers of base pairs between foot and track - {\it i.e.,} moving from right to left on the graph. Large jumps at 7-8 and 12-13 are due to the repairing of mismatches by the invader; these mismatches are present by design to prevent enzymatic cleavage of the track. \teo{Statistical errors are similar in magnitude to the size of symbols; a more detailed discussion is given in the Supplementary Information.}
}
\label{fig_motor_walker_FEP}
\end{figure}
The Turberfield group has also introduced a two-footed DNA walker that is designed to achieve directional motion without modifying its track, a continuous single strand consisting of multiple binding sites \cite{Bath2009}.  The walker is \teo{intended} to step in a foot-over-foot fashion; its full design is given in Ref. \cite{Bath2009}.  It is intended to exist either with a single foot bound to the track and the other raised, or with both feet bound to \teo{overlapping} adjacent sites, as illustrated in Fig.\,\ref{fig_walker}\,(a) and (b). The feet must \teo{overlap}, as \teo{competition} means that one or other of the feet will \teo{then} have a raised toehold domain.  \teo{A} single-stranded fuel can bind \teo{to this raised domain}, initiating displacement of the track  and raising the foot. \teo{The fuel is designed to selectively raise the `back' foot due to asymmetry within the system}.

An earlier oxDNA study \cite{Ouldridge_walker_2013} revealed that the walker had a tendency to `overstep', or bind to two non-adjacent sites (Fig.\,\ref{fig_walker}\,(c)). This is  disastrous for the walker, leaving both feet bound to the track without a free toehold to initiate foot-lifting. This tendency was observed even with tracks under moderate tension ($\sim 15$\,pN). Although tension could not prevent overstepping, increased disruption (fraying)  of the front foot/track duplex   hinted that recovery might be easier in the presence of tension, due to destabilisation of the overstepped state.

I now present evidence that tension does facilitate recovery from the overstepped state. I simulate toehold-free (blunt-ended) displacement of the track from the overstepped foot by a fuel molecule, \teo{as} illustrated in Fig.\,\ref{fig_walker}\,(d), both with and without a tension of 14.6\,pN applied to the track. Langevin Dynamics simulations \teo{enhanced by FFS} are used to estimate relative rates, and free-energy profiles of displacement are estimated using VMMC aided by US. Full methodological details are provided in the Supplementary Information.  Kinetic simulations indicate that displacement is $\sim 5 \times 10^3$ faster with tension than without, and the free-energy profiles (Fig.\,\ref{fig_motor_walker_FEP}) show that a difference of about $\Delta \Delta G_{\rm tension} \approx 6.4k_{\rm B}T$ develops between the two systems as displacement progresses. 

When bound to two non-adjacent sites \teo{as in Fig.\,\ref{fig_walker}\,(c)}, the walker tends to constrain the track and reduce its  possible extension. Importantly, as individual base pairs between the forward edge of the front foot and the track are disrupted, the track is able to extend incrementally further (see Fig.\,\ref{fig_walker}\,(d)). As a consequence, these base pairs break more easily than without tension -- they have a higher tendency to be frayed prior to fuel binding, providing an effective toehold for displacement. Further, because base pairs with the fuel are more stable than with the track, displacement of the track is biased towards success (as evidenced by the slope at the start of displacement in Fig.\,\ref{fig_motor_walker_FEP} for the system under tension). Detailed inspection of the FFS results indicates that faster displacement is due to both easier binding and a higher subsequent probability of displacement success. If the thermodynamic results were used to estimate relative rates via $\exp(-\Delta \Delta G_{\rm tension}/k_{\rm B}T)$, a value of approximately 600 would be obtained. The \teo{statistical errors  associated with the kinetic estimates in particular are} substantial, so it is possible that the difference is due to random error. However, it is also possible that the free-energy profile  fails to highlight subtle kinetic effects, such as those that were observed in conventional displacement. Regardless, the two results clearly predict a substantial increase in walker recovery rate of $\sim 10^3$ or more under the application of moderate tension.

\begin{figure}
\centering
\includegraphics[width=0.45\textwidth]{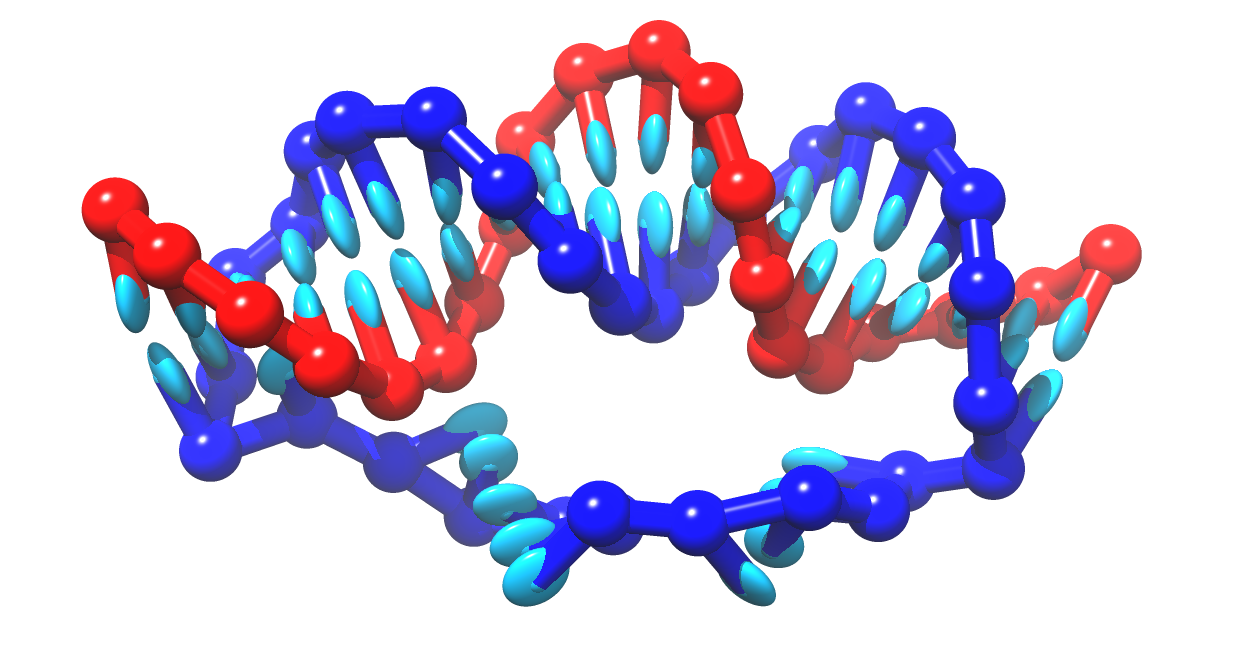}
\centering \caption{\footnotesize Generating tension that will favour displacement in a ``self contained" manner by using a short circular strand as the incumbent.
}
\label{fig_tension_disp}
\end{figure}

The fact that tension can aid walker recovery highlights the possibility of building contingency into nanodevice operation, thereby allowing for fail-safe behaviour. It also demonstrates the possibility of enhancing displacement through tension, an alternative way to modulate kinetics. In this case, tension is externally applied to the track; \teo{one possibility is to stretch the track between two points of an origami structure}. One could generate tension in a more self-contained way by using circular substrates, as shown in Fig. \ref{fig_tension_disp}; \teo{indeed,} similar constructs have been studied experimentally \cite{Hao2011,Wang2013}. \teo{Here, a sufficiently long duplex section will apply an extensional stress to a sufficiently short single-stranded section, and will experience a compression in response, leading to the bending shown in Fig. \ref{fig_tension_disp}.} Displacement rates in such a system could be modulated by secondary structure within, or binding to, the single-stranded loop of the substrate - this modulation would be freed from any constraints of the invader sequence. 

\section{Discussion}
DNA nanotechnology has enormous potential, but providing a solid theoretical underpinning is vital if it is to truly flourish. This understanding can  be achieved through a combination of careful experimentation and modelling. Certain groups are now taking up the challenge of meticulously understanding the \teo{processes} involved by carefully measuring important reactions in a variety of contexts \cite{Zhang_disp_2009,Sobczak2012,Myhrvold2013,Ke2012b,Martin2012,Tomov2013,Johnson-Buck2013,Tsukanov2013}. In this article I have discussed a coarse-grained model of DNA, oxDNA, that can complement these efforts.  OxDNA is an attempt to incorporate the thermodynamics of DNA as parameterised by SantaLucia's nearest-neighbour model into a structural and mechanical framework consistent with current knowledge. It can then be used to study the behaviour of `ideal' DNA, exploring the possibilities of certain motifs, testing whether new experimental observations are consistent with current understanding, and revealing which properties of DNA have an influence on certain phenomena.  

I have reviewed a number of applications of oxDNA, demonstrating that it provides quantitative understanding of aspects of displacement processes and can offer qualitative insight into reaction mechanisms and novel motifs. In each case, a physically reasonable underlying cause for the observation in question was identified, such as stress that is generated or relieved during displacement. These effects should therefore be manifest in real DNA, rather than simply being artefacts the model or simulation technique. Intriguing questions remain open, as there are many possible motifs available for study. For example, displacement reactions have recently been reported in the context of unconventional left-handed super-helices formed by conventional B-DNA domains in a complex pseudoknotted structure \cite{Li2014} -- such a system would an ideal candidate for study with oxDNA. Other possibilities include studying the influence of single-stranded hairpins, which are frequently present both by accident and design, on hybridisation and displacement kinetics, and considering general properties of displacement cascades on surfaces \cite{Teichmann2014}.

oxDNA's strengths are its efficiency and good overall representation of the structural, mechanical and thermodynamic changes associated with short duplex formation.   To achieve efficiency, however, the description of nucleotides is extremely simple and electrostatics in particular is treated in a basic fashion. Some properties, such as the opening of internal denaturation bubbles and the structures of systems that are more complex than simple duplexes, are yet to be fully tested against experimental data. Characterising and improving these limitations will be the subject of further work.

oxDNA is parameterised to experimental data on the basic properties of DNA, but experimental results are not unequivocal. In particular, the typical state, stacking propensity and conformation of single strands  are poorly understood (as discussed in Refs. \cite{Ouldridge2011, Ouldridge_thesis}), with some authors even finding evidence for structure formation when none would be expected from Watson-Crick base pairing \cite{Sikora2013}. Single strands have an enormous role in nanotechnological systems, not least as the initial state of the hybridisation process, which itself is not unambiguously understood. \teo{Particularly pertinent open questions relate to the temperature-dependence of hybridisation, as discussed in Ref. \cite{Ouldridge_binding_2013}, and its behaviour near a surface \cite{Johnson-Buck2013}}. Mechanical properties of duplexes, and whether DNA can be described as a wormlike-chain at short length scales, are still the subject of debate \cite{Wiggins2006,Vafabakhsh2012,Vologodskii2013,Mazur2014}. Resolving these fundamental issues would be extremely useful in nanotechnology and beyond. It is likely that, as further characterisation is performed, some of the assumptions inherent in the current understanding of DNA (and hence oxDNA) will have to be revisited.

In this article I have focussed on the application of oxDNA to small, active systems. oxDNA has also been used to study DNA in a range of other contexts, including large duplexes under applied stress \cite{Doye2013}, and self-assembly driven phase transitions \cite{Rovigatti2014,Rovigatti2014b}. A modified version parameterised to describe RNA has recently been developed \cite{Sulc2014}. Although some progress has  been made with improving the experimental protocols for constructing large structures \cite{Sobczak2012,Myhrvold2013,Ke2012b,Martin2012}, the assembly pathways and systematic ways to optimise them remain elusive. Modelling the assembly of larger structures with oxDNA is a current goal, but remains extremely challenging due to the large number of rare-event binding processes that must be observed. Further development and refinement of enhanced dynamic sampling techniques, such as FFS, would assist in this process. It may be that oxDNA can best be used to study smaller sub-structures, and that this information can then be incorporated into models at the level of secondary structure \cite{Dirks2007,Schaeffer_thesis} or whole binding domains \cite{Phillips2009,Arbona2013,Reinhardt2014}.

\makeatletter 
\renewcommand{\thefigure}{S\@arabic\c@figure} 
\renewcommand{\thetable}{S\@arabic\c@table} 
\renewcommand{\theequation}{S\@arabic\c@equation} 

\begin{appendix}
\section{Simulation methods}
\label{Simulation Methods}
Thermodynamic properties of model DNA are obtained by averaging over the Boltzmann distribution
\begin{equation}
\rho({\bf r}^N, {\bf p}^N, {\bf q}^N, {\bf L}^N)  \hspace{2mm}  \propto \hspace{2mm} \exp({-\beta \mathcal{H}({\bf r}^N, {\bf p}^N, {\bf q}^N, {\bf L}^N) }),
\label{boltzmann distribution}
\end{equation}
where $\mathcal{H}$ is the system Hamiltonian, ${\bf r}$ and ${\bf q}$ are positional and orientational degrees of freedom and ${\bf p}$ and ${\bf L}$ are linear and angular momenta. The momenta can 
be separately integrated analytically, and thus the relative probability of a configuration is given simply by  $\exp(-\beta V({\bf r}^N,{\bf q}^N))$. For oxDNA, thermodynamic averages are typically calculated using the  Virtual Move Monte Carlo (VMMC) algorithm.\cite{Whitelam2007,Whitelam2009} Dynamical studies reported in this work use a Langevin Dynamics (LD) algorithm for rigid bodies,\cite{Davidchack2009} which samples from the Boltzmann distribution and generates diffusive particle motion.

\subsection{VMMC}
\label{methods:VMMC}
VMMC\cite{Whitelam2007,Whitelam2009} is a cluster-based Monte Carlo technique that is particularly 
 effective for dilute systems with strong, directional interactions such as oxDNA. Results reported in this work use the variant introduced in the appendix of Ref. \onlinecite{Whitelam2009}. The algorithm involves attempting moves of particle clusters that are generated in a way that reflects potential energy gradients in the system, and accepting these attempted moves with a probability that ensures the system samples from the canonical ensemble. As clusters of strongly interacting particles move together, VMMC equilibrates oxDNA systems  much more efficiently than traditional Monte Carlo algorithms. 
 
`Seed' moves of a single particles are used to explore the energy gradients in a system and generate clusters. For all the VMMC simulations reported in this work, the seed moves were;
\begin{itemize}
\item rotation of a nucleotide about its backbone site, with the axis chosen from a uniform random distribution and the angle from a normal distribution with mean of zero and a standard deviation of 0.12 radians.
\item translation of a nucleotide with the direction chosen from a uniform random distribution and the distance from a normal distribution with mean of zero  and a standard deviation of 1.02\,\AA. 
\end{itemize}

\subsubsection{Umbrella Sampling}
\label{methods:umbrella}
Many processes (such as the binding of two strands) are slow to equilibrate, despite the simplicity of oxDNA and the efficiency of VMMC as a simulation technique. Slow equilibration often arises when the system has multiple competing local minima of free energy, separated by large free-energy barriers. 
An artificial biasing weight $W({\bf r}^N, {\bf q}^N)$ can be used to enhance equilibration  by flattening these barriers.\cite{Torrie1977} This procedure, known as Umbrella Sampling (US), involves choosing $W({\bf r}^N, {\bf q}^N)$ to favour the states of high free energy that must be crossed to pass between the free energy minima. The expectation of any variable $A$ can be extracted from these biased simulations via
\begin{equation}
\langle A \rangle = \frac{\langle A({\bf r}^N, {\bf q}^N)/W({\bf r}^N, {\bf q}^N) \rangle_W }{\langle 1/W({\bf r}^N, {\bf q}^N) \rangle_W}.
\end{equation}
Here, $\langle \rangle_W$ indicates sampling from the biased ensemble in which states are visited with a probability proportional to \linebreak $W({\bf r}^N, {\bf q}^N) \exp(-\beta V({\bf r}^N, {\bf q}^N))$. It is often helpful to define a (low-dimensional) order parameter $\psi({\bf r}^N, {\bf q}^N)$ that describes the process in question, and to then define the bias $W(\psi)$ using this order parameter rather than through the individual coordinates.

\subsubsection{The Weighted Histogram Analysis Method}
\label{methods:WHAM}
Even with Umbrella Sampling it can be challenging to sample all relevant states of a system within a single simulation. It is sometimes advantageous to split the simulations into overlapping `windows' of state space that can be separately equilibrated more easily. The data from the individual windows can be combined to give the statistical properties of the whole system by relating the windows using the overlapping regions. One systematic approach to achieve this is known as the `Weighted Histogram Analysis Method' (WHAM).\cite{Kumar1992} To use the WHAM algorithm, one must supply
\begin{itemize}
\item $ W^i (\psi)$, the biasing potential used in window $i$.
\item $ \rho^i(\psi)$, the normalised distribution measured during simulation window $i$.
\item $n^i$, the total number of VMMC steps taken over all simulations in window $i$.
\end{itemize}
The WHAM algorithm takes these inputs and iteratively estimates the overall distribution $\rho(\psi)$ and the dimensionless free-energy of window $i$ ($F_i$), via the equations
\begin{equation}
\begin{array}{c}
\rho^\prime(\psi) = \frac{\sum_i  \rho^i(\psi) n^i}{\sum_i  n^i \exp(F_i) W^i(\psi) },  \\
\\
\rho(\psi) = \frac{\rho^\prime(\psi)}{\sum_\psi \rho^\prime(\psi)},\\
\\ 
\exp(-F^i) = \sum_\psi W^i(\psi) \rho(\psi).\\
\end{array}
\end{equation}
From an initial (trivial) choice of $F_i$, the measured data can be used to iterate these equations until convergence, resulting in an overall estimate of the unbiased occupancy of individual states, $\rho(\psi)$.

\subsection{Langevin Dynamics}
\label{methods:LD}
LD allows Newton's equations of motion for the particles that are explicit in the model to be augmented with random and dissipative forces due to an implicit solvent. These forces are chosen so that particles move diffusively and the system samples from the Boltzmann distribution. The kinetic results reported in this work were obtained using the quaternion-based algorithm for rigid bodies of Davidchack {\it et al.}\cite{Davidchack2009} Other work on oxDNA has been performed with an Andersen-like thermostat.\cite{Russo2009}

To use the LD algorithm, a nucleotide mass (which is taken as $m=315.75$\,Da for all nucleotides) and a moment of inertia tensor  (we treat the nucleotides as spherical with a moment of inertia $31.586$\,Da\,nm$^2$) are required.\cite{Ouldridge_thesis} 
The drag forces experienced by a particle must also be related to its generalised momenta via a friction tensor. \teo{We ignore the complexity of hydrodynamic coupling between particles}. For simplicity, we treat each nucleotide's interaction with the solvent as spherically symmetric which means that only linear and rotational damping coefficients $\gamma$ and $\Gamma$ remain as independent quantities. The investigations reported in this paper use values of $\gamma = 0.59$\,ps$^{-1} $ and $\Gamma =1.76$\,ps$^{-1} $. These values result in  diffusion coefficients of $D_{\rm sim} = 1.91 \times 10^{-9}$\,m$^{2}$s$^{-1}$ for a 14 base-pair duplex, higher than experimental measurements of $D_{\rm exp} = 1.19 \times 10^{-10}$\,m$^{2}$s$^{-1}$.{\cite{Lapham1997}} Accelerating diffusion in this manner allows simulations to explore more complex processes for a given CPU time. Our group has previously shown\cite{Ouldridge_binding_2013} that increased friction coefficients have almost no qualitative consequences for hybridization pathways in oxDNA, only an overall reduction in the reaction rate.  LD Simulations in this work use an integration time step of 8.55\,fs, which has been previously shown to reproduce the energies and kinetics obtained with shorter time steps for oxDNA.\cite{Ouldridge_thesis}

In the studies reported in this work, I compare the measured rates of similar processes; \teo{typically, displacement rates as certain system parameters are changed}. For such comparisons, the effects of the choice of dynamical method should be approximately the same in each case, meaning that  relative rates should be reasonably insensitive to the details of the dynamical algorithm. Furthermore, the numerical results are made meaningful by identifying  the physically reasonable factors that lead to them, rather than just reporting the results in isolation.

\subsubsection{Forward flux sampling}
\label{methods:FFS}
`Brute force' LD simulations are often not efficient enough to sample rare transitions accurately. Forward Flux Sampling (FFS) can be used to enhance the calculation of the flux between two local free-energy minima, and improve sampling from the trajectories that link the two regions ({\it reactive trajectories}).\cite{Allen2005,Allen2009} Here I outline the general method; our specific implementation is detailed in Section \ref{implementation}.

Firstly I define the term `flux'. Let $A$ and $B$ be two non-overlapping regions of phase space. The flux from $A$ to $B$ is defined in the following way.
\begin{quotation}
For an infinitely long simulation, the flux of  trajectories from $A$ to $B$ is $\Phi_{AB} = N_{AB}/{(\tau f_{A})}$, where $N_{AB}$ is the number of times the simulation leaves $A$ and then reaches $B$ without first returning to $A$,  ${\tau}$ is the total simulation time and $f_{A}$ is the fraction of  simulation time during which the system visited state $A$ more recently than state $B$.
\end{quotation} 
The concept of  flux generalises a transition rate for processes that are not absolutely instantaneous.

To use FFS, a one-dimensional discrete order parameter $Q$ is required so that non-intersecting interfaces $\lambda^n_{n-1}$ exist between consecutive values of $Q$. The lowest value of $Q$, $Q=-2$, defines state $A$, and the highest value $Q=Q_{\rm max}$ state $B$. Initially, a simulation is performed in the vicinity of $Q=-2$, and the flux of trajectories from $Q=-2$ to $Q=0$ is measured (in effect, one counts the trajectories that cross  the surface $\lambda^0_{-1}$ for the first time since leaving $Q=-2$). I choose to define the lowest value of $Q$ as $Q=-2$ because the algorithm is distinct for values of $Q>0$.

The total flux of trajectories from $A$ to $B$ then follows as the flux across  $\lambda^0_{-1}$ from $Q=-2$, multiplied by the conditional probability that these trajectories subsequently cross the interface $\lambda_{Q_{\rm max}-1}^{Q_{\rm max}}$ to reach $Q = Q_{\rm max}$ instead of returning to $Q=-2$, $P(\lambda_{Q_{\rm max}-1}^{Q_{\rm max}} | \lambda_{-1}^0)$. This probability can itself be expressed as a product of the success probabilities associated with reaching the next interface (rather than returning to $Q=-2$) for each intermediate case, $P(\lambda_{Q-1}^Q | \lambda_{Q-2}^{Q-1})$:
\begin{equation}
P(\lambda_{Q_{\rm max}-1}^{Q_{\rm max}} | \lambda_{-1}^0) = \prod_{Q=1}^{Q_{\rm max}}  P(\lambda_{Q-1}^Q | \lambda_{Q-2}^{Q-1}).
\label{FFS_eqn}
\end{equation}

\begin{figure}
\centering
\includegraphics[width=0.47\textwidth]{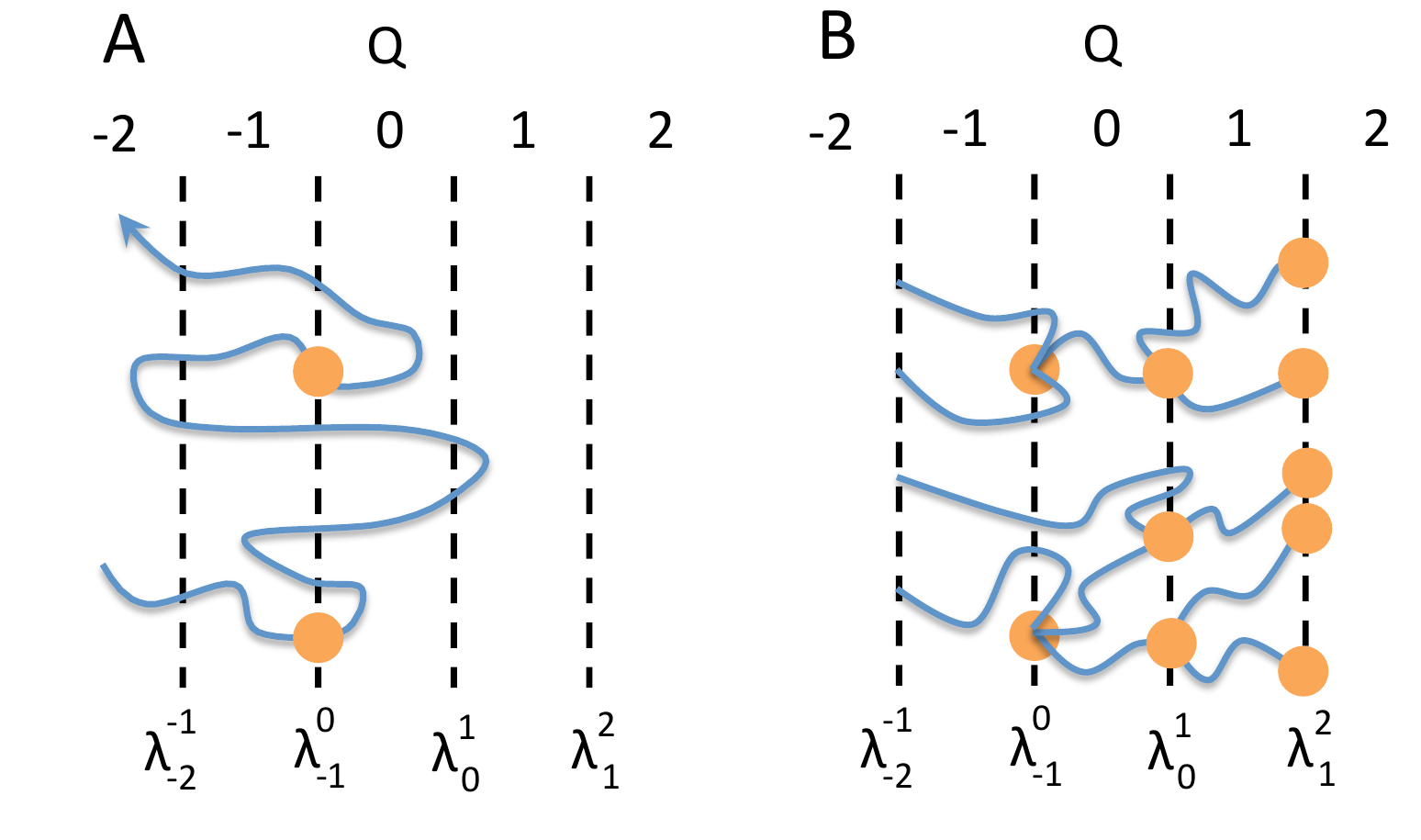}
\centering \caption{\footnotesize Schematic illustrations of FFS. An order parameter $Q$ is defined with interfaces $\lambda$ separating distinct value of $Q$. We are interested in measuring the flux from $Q=-2$ to $Q=2$ in this example. A) The initial measurement of flux across the interface $\lambda^0_{-1}$. Orange dots indicate the crossings that contribute to the flux, and also the states used to launch subsequent stages of simulation. B) Direct FFS involves randomly launching many trajectories from the interface $\lambda^0_{-1}$, and measuring the probability of reaching $\lambda^1_{0}$ before returning to $\lambda^{-1}_{-2}$. This procedure is then repeated for successive interfaces, resulting in branched trajectories. Reprinted by Permission from T. E. Ouldridge {\it et al.,} DNA hybridization kinetics: zippering, internal displacement and sequence dependence, Nucleic Acids Research 41:8886-8895. Copyright (2013) Oxford University Press.\cite{Ouldridge_binding_2013}
\label{fig_FFS}}
\end{figure}

In this work, the `direct' FFS approach was used  to calculate the product in Equation \ref{FFS_eqn}. In this method, a flux simulation is run to calculate the flux from $Q=-2$ to $Q=0$, and generate states at the $\lambda^0_{-1}$ interface. These states are used as starting points for estimating $P(\lambda_{0}^1 | \lambda_{-1}^{0})$. The process is then iterated for subsequent interfaces, using the successes from the previous stage as initial configurations for the next. The branched trajectories obtained sample from the distribution of reactive trajectories. The procedure is illustrated schematically in Fig. \ref{fig_FFS}. 

\begin{table}
\begin{center}
\begin{tabular}{c|  c}
Strand  & Sequence ($5^\prime$--$3^\prime$) \\
\hline
Substrate & GAC ATG GAG A{\it CG TAG}\\
& {\it GGT ATT GAA TGA GGG} \\
Incumbent & T{\it CC CTC ATT CAA TAC CCT ACG} \\
Invader &  {\it CC CTC ATT CAA TAC CCT ACG}\\
\end{tabular}\vspace{-0.1in}
\caption{\footnotesize Sequences used to simulate zero-toehold displacement. Italic regions indicate the displacement domain. \label{disp_sequences}}
\end{center}
\end{table}

\begin{table*}
\begin{center}
\begin{tabular}{c c}
\hline
Strand & Sequence (5$^\prime$ -- 3$^\prime$) \\
\hline 
Track (3 sites) & \hilighty{C}\hilightb{AGCATC}\hilighty{C\^{T}TCAGC\^TTC}\hilightb{AGCATC}\hilighty{C\^{T}TCAGC\^TTC}\hilightb{AGCATC}\hilighty{C\^TTCAGC\^TTC}\hilightb{AGCATC}\hilighty{C}\\
Walker 1 &   \hilightg{GTATTATCGTTAGTCT}tttt\hilightb{GATGCT}\hilighty{GA\^GGCTGA\^GG}\hilightb{GATGCT}\\
Walker  2 & \hilightg{AGACTAACGATAATAC}tttt\hilightb{GATGCT}\hilighty{GA\^GGCTGA\^GG}\hilightb{GATGCT} \\
Fuel  & \hilighty{CCTCAGCCTC}\hilightb{AGCATC}  \\
\hline
\end{tabular}
\caption{\footnotesize Sequences used for simulating the DNA walker. Complementary regions are highlighted in the same colour (corresponding to the colours used for the base sites in Fig. 9 of the main text). The toehold domains are highlighted in blue, and the longer binding domains in yellow. The duplex that holds the walker together is shown in green. Circumflexes indicate mismatched bases between the foot and track (which prevent the track from being cleaved by nicking enzymes in experiment, and also favour foot-lifting by the fuel).  Although the fuel can in principle bind to either toehold of both walker strands, only the 5$^\prime$ end will naturally lead to displacement.}
\label{walker_sequences}
\end{center}
\end{table*} 

\subsubsection{Error estimation}
\label{naive_errors}
In the initial work on FFS, the authors suggested calculating statistical error by treating each stage of the simulation as independent attempt to estimate a probability $P(\lambda_{i}^{i+1} | \lambda_{i-1}^{i})$ through a number of independent trials. Such an approach gives the variance in the measured value of $P(\lambda_{i}^{i+1} | \lambda_{i-1}^{i})$ as
\begin{equation}
\sigma_i^2 =   P(\lambda_{i}^{i+1} | \lambda_{i-1}^{i})(1-P(\lambda_{i}^{i+1} | \lambda_{i-1}^{i}))/N_i,
\end{equation}
where $N_i$ is the number of trials launched from interface $i$. The overall variance in the flux measurement would then follow by summing the individual variances from each stage. 

 \begin{table*}
\begin{center}
\begin{tabular}{c c c c c c}
Order parameter  &  Distance   &   Nearly-formed &Number of base pairs &Number of base pairs  & Distance      \\
$Q$ & $d$/nm &  base pairs $n$ & $A$ with $E<-2.98$\,kcal\,mol$^{-1}$& $B$ with $E<-2.98$\,kcal\,mol$^{-1}$& $d_2$/nm \\
\hline
\\
-2 & $d>5.11 $& $\sim$ & $\sim$ & $\sim$ & $\sim$\\
-1 & $5.11>d> 3.41$&  $\sim$ & $\sim$ & $\sim$ & $\sim$\\
0 & $3.41>d> 2.56$& $\sim$ & $\sim$ & $\sim$ & $\sim$\\
1 & $2.56>d>1.70 $& $\sim$ & $\sim$ & $\sim$ & $\sim$\\
2 & $1.70>d>0.852 $& $\sim$ & $\sim$ & $\sim$ & $\sim$\\
3 & $0.852>d $& 0 & 0 & $\sim$ & $\sim$\\
4 & $\sim$& 1 & 0 & $\sim$ & $\sim$\\
5 &$\sim$&  $\sim$ & $A=1$ &$\sim$ &  $\sim$\\
6 & $\sim$& $\sim$ &$\sim$  & $1<B<4$& $\sim$\\
7 & $\sim$& $\sim$ & $\sim$ & \multicolumn{2}{c}{$4 < B<20$ or ($B=20$ \& $d_2<2.56$)}\\
8 & $\sim$& $\sim$ & $\sim$ & $B=20$ & $d_2 \geq 2.56$  \\
\hline
\end{tabular}
\caption{\footnotesize Order parameter definition for FFS simulations of blunt-ended displacement. $d$ and $d_2 $ are minimum distances within certain sets of pairs of bases, measured as the separation of hydrogen-bonding sites of a pair. ``Nearly-formed" base pairs are defined in the text. The pairs of bases considered for $d$, $n$ and $A$ are only the 3 base pairs at either end of the intended invader/substrate duplex. $B$ includes {\it all} intended base pairs in the invader/substrate duplex. $d_2$ includes all base pairs in the original incumbent/substrate duplex. ``$\sim$" indicates that no additional restriction is placed on this collective degree of freedom that is not covered by other explicitly stated requirements.\label{blunt_end_FFS_op}}
\end{center}
\end{table*}

\begin{table}
\begin{center}
\begin{tabular}{c c}

Number of replicas & 10\\
\hline 
\\
Number of simulations per & 10  \\
replica for flux across $\lambda^0_{-1}$\\
\\
Initialisation time & 85.5\\
per simulation /ns\\
\\
Total crossings of $\lambda^0_{-1}$ & 10000  \\
(total time taken/$\mu$s) & 181  \\
\\
Average flux across $\lambda^0_{-1}$/$\mu$s$^{-1}$ & 55.2 \\
\hline
\\
Target interface & Total states loaded/successes \\
$\lambda^1_{0}$  & 100000/41957\\
$\lambda^2_{1}$  & 100000/42826\\
$\lambda^3_{2}$  & 100000/21010\\
$\lambda^4_{3}$  & 100000/ 14469\\
$\lambda^5_{4}$  &  800000/41467\\
$\lambda^6_{5}$  &  400000/12297\\
$\lambda^7_{6}$  &  200000/2797\\
$\lambda^8_{7}$  &  5000/267\\
\hline
\\
Overall flux to & 0.0149\\ 
full displacement/s$^{-1}$ & 0.026\\
(individual replica & 0.00482\\
 estimates) & 0.0356\\
 & 0.000975\\
 & 0.00362 \\
 & 0.0220 \\
 & 0.0200 \\
 & 0.00475 \\
& 0.139 \\
\hline
\end{tabular}
\end{center}
\caption{Simulation details and data obtained from FFS simulations of blunt-ended displacement.\label{blunt_end_FFS_protocols}}
\end{table}

Our group followed this approach in Refs. \onlinecite{Ouldridge_walker_2013} and \onlinecite{Srinivas2013}. However, this method underestimates the true variability of the final result, as it assumes that the initial configurations at each interface are a truly representative ensemble. In some cases, particularly when bottlenecks appear in the process, this assumption can be poorly justified. For example, analysis of the simulation results for the zero-toehold displacement presented in Ref. \onlinecite{Srinivas2013} showed that only a few of the initial configurations obtained at the $\lambda^0_{-1}$ interface subsequently spawned trajectories that led to complete displacement, raising the possibility that the true uncertainty in the result is much higher than suggested.

An alternative approach is to estimate statistical errors by running a number of completely separate FFS protocols and comparing the independent estimates of the overall flux. I have used this procedure for the simulations of walker recovery presented in this work. To check the reliability of our earlier result on TMSD, I also present data for zero-toehold displacement obtained using this approach in the main text. Overall, although error estimates may have been too small, the physical conclusions drawn from earlier work are robust.

\section{Systems}
\subsection{Zero-toehold displacement}

Simulations of zero-toehold displacement used the same setup as Ref. \onlinecite{Srinivas2013}. Simulations involved one invader, one substrate and one incumbent  in a periodic cubic cell of $1.67\times 10^{-20}$~l \teo{(litres)}, at a temperature of $T=298.15$\,K. Table \ref{disp_sequences} contains strand sequences. As in the original work,\cite{Srinivas2013} the sequence-independent version of oxDNA was used, and only base pairs native to the initial incumbent/substrate and final invader/substrate duplexes were assigned non-zero binding strength. This simplification eliminates the metastable misbonded configurations, which can hamper FFS without significantly affecting system behaviour.\cite{Srinivas2013} The sequence-independent parameterisation of oxDNA can simplifies results as it shows only the generic behaviour of DNA, rather than sequence-specific effects.

\section{Simulation protocols}
\label{implementation}
 \begin{table*}
\begin{center}
\begin{tabular}{c c c c c }
Order parameter  &  Distance   &   Nearly-formed &Number of base pairs  &  Number of base pairs     \\
$Q$ & $d$/nm &  base pairs $n$ & $A$ with $E<E_A$& $B$ with $E<E_B$ \\
\hline
\\
-2 & $d> 6.81 $& $\sim$ & $\sim$ & $\sim$\\
-1& $6.81>d>5.11 $& $\sim$ & $\sim$ & $\sim$\\
0& $5.11>d> 4.26$&  $\sim$ & $\sim$ & $\sim$\\
1 & $4.26>d> 3.41$& $\sim$ & $\sim$ & $\sim$\\
2 & $3.41>d> 2.56$& $\sim$ & $\sim$ & $\sim$\\
3 & $2.56>d>1.70 $& $\sim$ & $\sim$ & $\sim$\\
4 & $1.70>d>0.852 $& $\sim$ & $\sim$ & $\sim$\\
5 & $0.852>d $& 0 & 0 & 0\\
6 & $\sim$& 1 & 0 & 0\\
7 &$\sim$&  $\sim$ & \multicolumn{2}{c}{($A \geq 1$\, \& $B=0$) {\rm or} ($A = 1$\, \& $B=1$ ) }  \\
8 & $\sim$& $\sim$ &  \multicolumn{2}{c}{($A \geq 2$\, \& $B=1$) {\rm or} ($A = 2$\, \& $B=2$ ) }  \\
9 & $\sim$& $\sim$ & $A \geq 3$ &$4 > B\geq 2$\\
10 & $\sim$& $\sim$ & $\sim$ &$16 > B\geq 4$ \\
11 & $\sim$& $\sim$ & $\sim$ & 16\\
\hline
\end{tabular}
\caption{\footnotesize Order parameter definition for FFS simulations of walker recovery by fuel displacement. $d$ is the minimum distance between bases that are intended to form base pairs in the final invader/substrate duplex, measured as the separation of hydrogen-bonding sites. ``Nearly-formed" base pairs are defined in the text. All base-pairing contributions to the order parameter include only those base pairs that are intended to be present in the final invader/substrate duplex. $E_A=-1.43$\,kcal\,mol$^{-1}$ and $E_B=-1.79$\,kcal\,mol$^{-1}$. ``$\sim$" indicates that no additional restriction is placed on this collective degree of freedom that is not covered by other explicitly stated requirements. \label{recovery_FFS_OP}}
\end{center}
\end{table*}

\begin{table}
\begin{center}
\begin{tabular}{c c c }
System & With tension & Without tension \\
\hline 
\\
Number of replicas & 10 & 10\\
\hline 
\\
Number of simulations per & 10 & 10   \\
replica for flux across $\lambda^0_{-1}$\\
\\
Initialisation time & 2.565 & 2.565 \\
per simulation /$\mu$s\\
\\
Total crossings of $\lambda^0_{-1}$ & 2557 & 3090  \\
(total time taken/$\mu$s) & 94.2 & 126 \\
\\
Average flux across $\lambda^0_{-1}$/$\mu$s$^{-1}$ & 27.1 & 24.5 \\
\hline
\\
Target interface & \multicolumn{2}{c}{Total states loaded/successes} \\
$\lambda^1_{0}$  &40000/18156 & 40000/17850\\
$\lambda^2_{1}$  &39803/21181 & 38975/20341\\
$\lambda^3_{2}$  &37600/21154 & 39947/22017\\
$\lambda^4_{3}$  & 40000/22095 & 40000/21060 \\
$\lambda^5_{4}$  &71200/20636  & 76000/18448 \\
$\lambda^6_{5}$  &40000/9708  & 80000/11296 \\
$\lambda^7_{6}$  &99000/3353  &200000/1744  \\
$\lambda^8_{7}$  &37939/5869  & 200000/5878 \\
$\lambda^9_{8}$  &20000/3981  & 160000/2629 \\
$\lambda^{10}_{9}$  & 5000/2137 & 30000/5105 \\
$\lambda^{11}_{10}$  & 1880/909 & 3880/773 \\
\hline
\\
Overall flux to & 135 &	0.00438\\ 
full displacement/s$^{-1}$ & 37.1&	0.00340\\
(individual replica & 45.5 &	0.0332\\
 estimates) & 32.3 &	0.00131\\
 &	58.1 &	0.00384	\\
&	7.96 &	0.00492	\\
&	1.49 &	0.00987	\\
&	47.9 &	0.01345	\\
&	18.3 &	0.00376	\\
&	11.3 &	0.00167	\\
\hline
\end{tabular}
\end{center}
\caption{Simulation details and data obtained from FFS simulations of walker recovery.\label{recovery_FFS_protocols}}
\end{table}

\subsection{Overstepped walker recovery}
Walker simulations involved the two strands that constitute the walker itself, a track containing three binding sites \teo{with one extra base at either end to limit the effects of truncating the track,} and a fuel strand  in a periodic cubic cell of $7.73\times 10^{-20}$~l.  Table \ref{walker_sequences} displays the sequences. A temperature of 310\,K was used, close to experimental conditions.\cite{Bath2009} For simulations involving tension, a constant force of 14.6\,pN was applied to the backbone sites at either end of the track. Consistent with earlier work on the walker using oxDNA,\cite{Ouldridge_walker_2013} the sequence-independent parameterisation of oxDNA was used. To simplify sampling, only expected interstrand bonds ({\it i.e.,} those intended in the design) were assigned non-zero binding strength, although any intrastrand base-pairing was allowed.

\begin{table*}
\begin{center}
\begin{tabular}{c | c c c c | c c c c c c c c c c c c c c c c}
&\multicolumn{4}{c |}{$\psi_2=0$, $\psi_3=$} & \multicolumn{16}{c}{$\psi_3=3$, $\psi_2=$}\\
$\psi_1$ & 0 & 1 & 2 & 3 & 1 & 2 & 3 & 4 & 5 & 6 & 7 & 8 & 9 & 10 & 11 & 12 & 13 & 14 & 15 & 16 \\
\hline
0 & 0 & 0 & 0 & 0 & 0 & 0 & 0 & 0 & 0 & 0 & 0 & 0 & 0 & 0 & 0 & 0 & 0 & 0 & 0 & 0\\
1 & 1 & 20 & 500 & 4000 & $1.2 \times 10^8$ & $1.2 \times 10^8$ & $1.2 \times 10^8$ &0 & 0 & 0 & 0 &0 & 0 & 0 & 0 & 0 & 0 & 0 & 0 & 0\\
2 & 1 & 20 & 500 & 4000 & $1.2 \times 10^8$ & $1.2 \times 10^8$ & $1.2 \times 10^8$ &0 & 0 & 0 & 0 &0 & 0 & 0 & 0 & 0 & 0 & 0 & 0 & 0\\
3 & 1 & 20 & 500 & 4000 & $1.2 \times 10^8$ & $1.2 \times 10^8$ & $1.2 \times 10^8$ &0 & 0 & 0 & 0 &0 & 0 & 0 & 0 & 0 & 0 & 0 & 0 & 0\\ 
4 & 1 & 20 & 500 & 4000 & $1.2 \times 10^8$ & $1.2 \times 10^8$ & $1.2 \times 10^8$ &0 & 0 & 0 & 0 &0 & 0 & 0 & 0 & 0 & 0 & 0 & 0 & 0\\
5 & 1 & 20 & 500 & 4000 & $1.2 \times 10^8$ & $1.2 \times 10^8$ & $1.2 \times 10^8$ &0 & 0 & 0 & 0 &0 & 0 & 0 & 0 & 0 & 0 & 0 & 0 & 0\\
6 & 1 & 20 & 500 & 4000 & $1.2 \times 10^8$ & $1.2 \times 10^8$ & $1.2 \times 10^8$ &0 & 0 & 0 & 0 &0 & 0 & 0 & 0 & 0 & 0 & 0 & 0 & 0\\
7 & 1 & 20 & 500 & 4000 & $1.2 \times 10^8$ & $1.2 \times 10^8$ & $1.2 \times 10^8$ &0 & 0 & 0 & 0 &0 & 0 & 0 & 0 & 0 & 0 & 0 & 0 & 0\\
8 & 1 & 20 & 500 & 4000 & $1.2 \times 10^8$ & $1.2 \times 10^8$ & $1.2 \times 10^8$ &0 & 0 & 0 & 0 &0 & 0 & 0 & 0 & 0 & 0 & 0 & 0 & 0\\
9 & 1 & 20 & 500 & 4000 & $1.2 \times 10^8$ & $1.2 \times 10^8$ & $1.2 \times 10^8$ &0 & 0 & 0 & 0 &0 & 0 & 0 & 0 & 0 & 0 & 0 & 0 & 0\\
10 & 1 & 20 & 500 & 4000 & $1.2 \times 10^8$ & $1.2 \times 10^8$ & $1.2 \times 10^8$ &0 & 0 & 0 & 0 &0 & 0 & 0 & 0 & 0 & 0 & 0 & 0 & 0\\
11 & 1 & 20 & 500 & 4000 & $1.2 \times 10^8$ & $1.2 \times 10^8$ & $1.2 \times 10^8$ &0 & 0 & 0 & 0 &0 & 0 & 0 & 0 & 0 & 0 & 0 & 0 & 0\\
12 & 1 & 20 & 500 & 4000 & $1.2 \times 10^8$ & $1.2 \times 10^8$ & $1.2 \times 10^8$ &0 & 0 & 0 & 0 &0 & 0 & 0 & 0 & 0 & 0 & 0 & 0 & 0\\
13 & 1 & 20 & 500 & 4000 & $1.2 \times 10^8$ & $1.2 \times 10^8$ & $1.2 \times 10^8$ &0 & 0 & 0 & 0 &0 & 0 & 0 & 0 & 0 & 0 & 0 & 0 & 0\\
14 & 1 & 20 & 500 & 4000 & $1.2 \times 10^8$ & $1.2 \times 10^8$ & $1.2 \times 10^8$ &0 & 0 & 0 & 0 &0 & 0 & 0 & 0 & 0 & 0 & 0 & 0 & 0\\
15 & 1 & 20 & 500 & 4000 & $1.2 \times 10^8$ & $1.2 \times 10^8$ & $1.2 \times 10^8$ &0 & 0 & 0 & 0 &0 & 0 & 0 & 0 & 0 & 0 & 0 & 0 & 0\\
16 & 1 & 20 & 500 & 4000 & $1.2 \times 10^8$ & $1.2 \times 10^8$ & $1.2 \times 10^8$ &0 & 0 & 0 & 0 &0 & 0 & 0 & 0 & 0 & 0 & 0 & 0 & 0\\
17 & 1 & 20 & 500 & 4000 & $1.2 \times 10^8$ & $1.2 \times 10^8$ & $1.2 \times 10^8$ &0 & 0 & 0 & 0 &0 & 0 & 0 & 0 & 0 & 0 & 0 & 0 & 0\\
18 & 1 & 20 & 500 & 4000 & $1.2 \times 10^8$ & $1.2 \times 10^8$ & $1.2 \times 10^8$ &0 & 0 & 0 & 0 &0 & 0 & 0 & 0 & 0 & 0 & 0 & 0 & 0\\
19 & 1 & 20 & 500 & 4000 & $1.2 \times 10^8$ & $1.2 \times 10^8$ & $1.2 \times 10^8$ &0 & 0 & 0 & 0 &0 & 0 & 0 & 0 & 0 & 0 & 0 & 0 & 0\\
20 & 1 & 20 & 500 & 4000 & $1.2 \times 10^8$ & $1.2 \times 10^8$ & $1.2 \times 10^8$ &0 & 0 & 0 & 0 &0 & 0 & 0 & 0 & 0 & 0 & 0 & 0 & 0\\
\end{tabular}
\caption{\footnotesize Biasing potential $W(\psi)$ for window 1 of an overstepped walker without tension, with $\psi = (\psi_1, \psi_2, \psi_3)$ a 3-dimensional order parameter. Due to their definitions, states with $\psi_2>0$ and $\psi_3 \neq 3$ are impossible.   \label{recovery_US_OP1_noT}}
\end{center}
\end{table*}

 \begin{table*}
\begin{center}
\begin{tabular}{c | c c c c | c c c c c c c c c c c c c c c c}
&\multicolumn{4}{c |}{$\psi_2=0$, $\psi_3=$} & \multicolumn{16}{c}{$\psi_3=3$, $\psi_2=$}\\
$\psi_1$ & 0 & 1 & 2 & 3 & 1 & 2 & 3 & 4 & 5 & 6 & 7 & 8 & 9 & 10 & 11 & 12 & 13 & 14 & 15 & 16 \\
\hline
0 & 0 & 0 & 0 & 0 & 0 & 0 & 0 & 0 & 0 & 0 & 0 & 0 & 0 & 0 & 0 & 0 & 0 & 0 & 0 & 0\\
1 & 0 & 0 & 0 & 0 & 0 & 0 & 0 & 0 & 0 & 0 & 0 & 0 & 0 & 0 & 0 & 0 & 0 & 0 & 0 & 0\\
2 & 0 & 0 & 0 & 0 & 0 & 0 & 0 & 0 & 0 & 0 & 0 & 0 & 0 & 0 & 0 & 0 & 0 & 0 & 0 & 0\\
3 & 0 & 0 & 0 & 0 & 0 & 0 & 0 & 0 & 0 & 0 & 0 & 0 & 0 & 0 & 0 & 0 & 0 & 0 & 0 & 0\\
4 & 0 & 0 & 0 & 0 & 0 & 0 & 0 & 0 & 0 & 0 & 0 & 0 & 0 & 0 & 0 & 0 & 0 & 0 & 0 & 0\\
5 & 0 & 0 & 0 & 0 & 0 & 0 & 0 & 0 & 0 & 0 & 0 & 0 & 0 & 0 & 0 & 0 & 0 & 0 & 0 & 0\\
6 & 0 & 0 & 0 & 0 & 0 & 0 & 0 & 0 & 0 & 0 & 0 & 0 & 0 & 0 & 0 & 0 & 0 & 0 & 0 & 0\\
7 & 0 & 0 & 0 & 0 & 0 & 0 & 0 & 0 & 0 & 0 & 0 & 0 & 0 & 0 & 0 & 0 & 0 & 0 & 0 & 0\\
8 & 0 & 0 & 0 & 0 & 0 & 0 & 0 & 0 & 0 & 0 & 0 & 0 & 0 & 0 & 0 & 0 & 0 & 0 & 0 & 0\\
9 & 0 & 0 & 0 & 0 & 0 & 0 & 0 & 0 & 0 & 0 & 0 & 0 & 0 & 0 & 0 & 0 & 0 & 0 & 0 & 0\\
10 & 0 & 0 & 0 & 0 & 0 & 0 & 0 & 0 & 0 & 0 & 0 & 0 & 0 & 0 & 0 & 0 & 0 & 0 & 0 & 0\\
11 & 0 & 0 & 0 & 0 & 1 & 1 & 1 & 1 & 1 & 1 & 1 & 1 & 1 & 1 & 1 & 1 & 1 & 1 & 1 & 1\\
12 & 0 & 0 & 0 & 0 & 1 & 1 & 1 & 1 & 1 & 1 & 1 & 1 & 1 & 1 & 1 & 1 & 1 & 1 & 1 & 1\\
13 & 0 & 0 & 0 & 0 & 100 & 100 & 100 & 100 & 100 & 100 & 100 & 100 & 100 & 100 & 100 & 100 & 100 & 100 & 100 & 100\\
14 & 0 & 0 & 0 & 0 & 100 & 100 & 100 & 100 & 100 & 100 & 100 & 100 & 100 & 100 & 100 & 100 & 100 & 100 & 100 & 100\\
15 & 0 & 0 & 0 & 0 & 100 & 100 & 100 & 100 & 100 & 100 & 100 & 100 & 100 & 100 & 100 & 100 & 100 & 100 & 100 & 100\\
16 & 0 & 0 & 0 & 0 & 100 & 100 & 100 & 100 & 100 & 100 & 100 & 100 & 100 & 100 & 100 & 100 & 100 & 100 & 100 & 100\\
17 & 0 & 0 & 0 & 0 & 100 & 100 & 100 & 100 & 100 & 100 & 100 & 100 & 100 & 100 & 100 & 100 & 100 & 100 & 100 & 100\\
18 & 0 & 0 & 0 & 0 & 100 & 100 & 100 & 100 & 100 & 100 & 100 & 100 & 100 & 100 & 100 & 100 & 100 & 100 & 100 & 100\\
19 & 0 & 0 & 0 & 0 & 100 & 100 & 100 & 100 & 100 & 100 & 100 & 100 & 100 & 100 & 100 & 100 & 100 & 100 & 100 & 100\\
20 & 0 & 0 & 0 & 0 & 100 & 100 & 100 & 100 & 100 & 100 & 100 & 100 & 100 & 100 & 100 & 100 & 100 & 100 & 100 & 100\\
\end{tabular}
\caption{\footnotesize Biasing potential $W(\psi)$ for window 2 of an overstepped walker without tension, with $\psi = (\psi_1, \psi_2, \psi_3)$ a 3-dimensional order parameter. Due to their definitions, states with $\psi_2>0$ and $\psi_3 \neq 3$ are impossible. \label{recovery_US_OP2_noT}}
\end{center}
\end{table*}

 \begin{table*}
\begin{center}
\begin{tabular}{c | c c c c | c c c c c c c c c c c c c c c c}
&\multicolumn{4}{c |}{$\psi_2=0$, $\psi_3=$} & \multicolumn{16}{c}{$\psi_3=3$, $\psi_2=$}\\
$\psi_1$ & 0 & 1 & 2 & 3 & 1 & 2 & 3 & 4 & 5 & 6 & 7 & 8 & 9 & 10 & 11 & 12 & 13 & 14 & 15 & 16 \\
\hline
0 & 0 & 0 & 0 & 0 & 0 & 0 & 0 & 0 & 0 & 0 & 0 & 0 & 0 & 0 & 0 & 0 & 0 & 0 & 0 & 0\\
1 & 0 & 0 & 0 & 0 & 1 & 1 & 1 & 1 & 1 & 1 & 1 & 1 & 1 & 1 & 1 & 1 & 1 & 1 & 1 & 1\\
2 & 0 & 0 & 0 & 0 & 1 & 1 & 1 & 1 & 1 & 1 & 1 & 1 & 1 & 1 & 1 & 1 & 1 & 1 & 1 & 1\\
3 & 0 & 0 & 0 & 0 & 1 & 1 & 1 & 1 & 1 & 1 & 1 & 1 & 1 & 1 & 1 & 1 & 1 & 1 & 1 & 1\\
4 & 0 & 0 & 0 & 0 & 1 & 1 & 1 & 1 & 1 & 1 & 1 & 1 & 1 & 1 & 1 & 1 & 1 & 1 & 1 & 1\\
5 & 0 & 0 & 0 & 0 & 1 & 1 & 1 & 1 & 1 & 1 & 1 & 1 & 1 & 1 & 1 & 1 & 1 & 1 & 1 & 1\\
6 & 0 & 0 & 0 & 0 & 1 & 1 & 1 & 1 & 1 & 1 & 1 & 1 & 1 & 1 & 1 & 1 & 1 & 1 & 1 & 1\\
7 & 0 & 0 & 0 & 0 & 2 & 2 & 2 & 2 & 2 & 2 & 2 & 2 & 2 & 2 & 2 & 2 & 2 & 2 & 2 & 2\\
8 & 0 & 0 & 0 & 0 &  400 & 400 & 400 & 400 & 400 & 400 & 400 & 400 & 400 & 400 & 400 & 400 & 400 & 400 & 400 & 400\\
9 & 0 & 0 & 0 & 0 & 400 & 400 & 400 & 400 & 400 & 400 & 400 & 400 & 400 & 400 & 400 & 400 & 400 & 400 & 400 & 400\\
10 & 0 & 0 & 0 & 0 &  400 & 400 & 400 & 400 & 400 & 400 & 400 & 400 & 400 & 400 & 400 & 400 & 400 & 400 & 400 & 400\\
11 & 0 & 0 & 0 & 0 &  400 & 400 & 400 & 400 & 400 & 400 & 400 & 400 & 400 & 400 & 400 & 400 & 400 & 400 & 400 & 400\\
12 & 0 & 0 & 0 & 0 & 400 & 400 & 400 & 400 & 400 & 400 & 400 & 400 & 400 & 400 & 400 & 400 & 400 & 400 & 400 & 400\\
13 & 0 & 0 & 0 & 0 & 0 & 0 & 0 & 0 & 0 & 0 & 0 & 0 & 0 & 0 & 0 & 0 & 0 & 0 & 0 & 0\\
14 & 0 & 0 & 0 & 0 & 0 & 0 & 0 & 0 & 0 & 0 & 0 & 0 & 0 & 0 & 0 & 0 & 0 & 0 & 0 & 0\\
15 & 0 & 0 & 0 & 0 & 0 & 0 & 0 & 0 & 0 & 0 & 0 & 0 & 0 & 0 & 0 & 0 & 0 & 0 & 0 & 0\\
16 & 0 & 0 & 0 & 0 & 0 & 0 & 0 & 0 & 0 & 0 & 0 & 0 & 0 & 0 & 0 & 0 & 0 & 0 & 0 & 0\\
17 & 0 & 0 & 0 & 0 & 0 & 0 & 0 & 0 & 0 & 0 & 0 & 0 & 0 & 0 & 0 & 0 & 0 & 0 & 0 & 0\\
18 & 0 & 0 & 0 & 0 & 0 & 0 & 0 & 0 & 0 & 0 & 0 & 0 & 0 & 0 & 0 & 0 & 0 & 0 & 0 & 0\\
19 & 0 & 0 & 0 & 0 & 0 & 0 & 0 & 0 & 0 & 0 & 0 & 0 & 0 & 0 & 0 & 0 & 0 & 0 & 0 & 0\\
20 & 0 & 0 & 0 & 0 & 0 & 0 & 0 & 0 & 0 & 0 & 0 & 0 & 0 & 0 & 0 & 0 & 0 & 0 & 0 & 0\\
\end{tabular}
\caption{\footnotesize Biasing potential $W(\psi)$ for window 3 of an overstepped walker without tension, with $\psi = (\psi_1, \psi_2, \psi_3)$ a 3-dimensional order parameter. Due to their definitions, states with $\psi_2>0$ and $\psi_3 \neq 3$ are impossible. \label{recovery_US_OP3_noT}}
\end{center}
\end{table*}

 \begin{table*}
\begin{center}
\begin{tabular}{c | c c c c | c c c c c c c c c c c c c c c c}
&\multicolumn{4}{c |}{$\psi_2=0$, $\psi_3=$} & \multicolumn{16}{c}{$\psi_3=3$, $\psi_2=$}\\
$\psi_1$ & 0 & 1 & 2 & 3 & 1 & 2 & 3 & 4 & 5 & 6 & 7 & 8 & 9 & 10 & 11 & 12 & 13 & 14 & 15 & 16 \\
\hline
0 & 0 & 0 & 0 & 0 & 0 & 0 & 0 & 0 & 0 & 0 & 0 & 0 & 0 & 0 & 0 & 0 & 0 & 0 & 0 & 0\\
1 & 1 & 10 & 100 & 250 & $2.5 \times 10^6$ & $1 \times 10^6$ & $5 \times 10^5$ &0 & 0 & 0 & 0 &0 & 0 & 0 & 0 & 0 & 0 & 0 & 0 & 0\\
2 & 1 & 10 & 100 & 250 & $2.5 \times 10^6$ & $1 \times 10^6$ & $5 \times 10^5$ &0 & 0 & 0 & 0 &0 & 0 & 0 & 0 & 0 & 0 & 0 & 0 & 0\\
3 & 1 & 10 & 100 & 250 & $2.5 \times 10^6$ & $1 \times 10^6$ & $5 \times 10^5$ &0 & 0 & 0 & 0 &0 & 0 & 0 & 0 & 0 & 0 & 0 & 0 & 0\\ 
4 & 1 & 10 & 100 & 250 & $2.5 \times 10^6$ & $1 \times 10^6$ & $5 \times 10^5$ &0 & 0 & 0 & 0 &0 & 0 & 0 & 0 & 0 & 0 & 0 & 0 & 0\\
5 & 1 & 10 & 100 & 250 & $2.5 \times 10^6$ & $1 \times 10^6$ & $5 \times 10^5$ &0 & 0 & 0 & 0 &0 & 0 & 0 & 0 & 0 & 0 & 0 & 0 & 0\\
6 & 1 & 10 & 100 & 250 & $2.5 \times 10^6$ & $1 \times 10^6$ & $5 \times 10^5$ &0 & 0 & 0 & 0 &0 & 0 & 0 & 0 & 0 & 0 & 0 & 0 & 0\\
7 & 1 & 10 & 100 & 250 & $2.5 \times 10^6$ & $1 \times 10^6$ & $5 \times 10^5$ &0 & 0 & 0 & 0 &0 & 0 & 0 & 0 & 0 & 0 & 0 & 0 & 0\\
8 & 1 & 10 & 100 & 250 & $2.5 \times 10^6$ & $1 \times 10^6$ & $5 \times 10^5$ &0 & 0 & 0 & 0 &0 & 0 & 0 & 0 & 0 & 0 & 0 & 0 & 0\\
9 & 1 & 10 & 100 & 250 & $2.5 \times 10^6$ & $1 \times 10^6$ & $5 \times 10^5$ &0 & 0 & 0 & 0 &0 & 0 & 0 & 0 & 0 & 0 & 0 & 0 & 0\\
10 & 1 & 10 & 100 & 250 & $2.5 \times 10^6$ & $1 \times 10^6$ & $5 \times 10^5$ &0 & 0 & 0 & 0 &0 & 0 & 0 & 0 & 0 & 0 & 0 & 0 & 0\\
11 & 1 & 10 & 100 & 250 & $2.5 \times 10^6$ & $1 \times 10^6$ & $5 \times 10^5$ &0 & 0 & 0 & 0 &0 & 0 & 0 & 0 & 0 & 0 & 0 & 0 & 0\\
12 & 1 & 10 & 100 & 250 & $2.5 \times 10^6$ & $1 \times 10^6$ & $5 \times 10^5$ &0 & 0 & 0 & 0 &0 & 0 & 0 & 0 & 0 & 0 & 0 & 0 & 0\\
13 & 1 & 10 & 100 & 250 & $2.5 \times 10^6$ & $1 \times 10^6$ & $5 \times 10^5$ &0 & 0 & 0 & 0 &0 & 0 & 0 & 0 & 0 & 0 & 0 & 0 & 0\\
14 & 1 & 10 & 100 & 250 & $2.5 \times 10^6$ & $1 \times 10^6$ & $5 \times 10^5$ &0 & 0 & 0 & 0 &0 & 0 & 0 & 0 & 0 & 0 & 0 & 0 & 0\\
15 & 1 & 10 & 100 & 250 & $2.5 \times 10^6$ & $1 \times 10^6$ & $5 \times 10^5$ &0 & 0 & 0 & 0 &0 & 0 & 0 & 0 & 0 & 0 & 0 & 0 & 0\\
16 & 1 & 10 & 100 & 250 & $2.5 \times 10^6$ & $1 \times 10^6$ & $5 \times 10^5$ &0 & 0 & 0 & 0 &0 & 0 & 0 & 0 & 0 & 0 & 0 & 0 & 0\\
17 & 1 & 10 & 100 & 250 & $2.5 \times 10^6$ & $1 \times 10^6$ & $5 \times 10^5$ &0 & 0 & 0 & 0 &0 & 0 & 0 & 0 & 0 & 0 & 0 & 0 & 0\\
18 & 1 & 10 & 100 & 250 & $2.5 \times 10^6$ & $1 \times 10^6$ & $5 \times 10^5$ &0 & 0 & 0 & 0 &0 & 0 & 0 & 0 & 0 & 0 & 0 & 0 & 0\\
19 & 1 & 10 & 100 & 250 & $2.5 \times 10^6$ & $1 \times 10^6$ & $5 \times 10^5$ &0 & 0 & 0 & 0 &0 & 0 & 0 & 0 & 0 & 0 & 0 & 0 & 0\\
20 & 1 & 10 & 100 & 250 & $2.5 \times 10^6$ & $1 \times 10^6$ & $5 \times 10^5$ &0 & 0 & 0 & 0 &0 & 0 & 0 & 0 & 0 & 0 & 0 & 0 & 0\\
\end{tabular}
\caption{\footnotesize Biasing potential $W(\psi)$ for window 1 of an overstepped walker with tension, with $\psi = (\psi_1, \psi_2, \psi_3)$ a 3-dimensional order parameter. Due to their definitions, states with $\psi_2>0$ and $\psi_3 \neq 3$ are impossible. \label{recovery_US_OP1_T}}
\end{center}
\end{table*}

 \begin{table*}
\begin{center}
\begin{tabular}{c | c c c c | c c c c c c c c c c c c c c c c}
&\multicolumn{4}{c |}{$\psi_2=0$, $\psi_3=$} & \multicolumn{16}{c}{$\psi_3=3$, $\psi_2=$}\\
$\psi_1$ & 0 & 1 & 2 & 3 & 1 & 2 & 3 & 4 & 5 & 6 & 7 & 8 & 9 & 10 & 11 & 12 & 13 & 14 & 15 & 16 \\
\hline
0 & 0 & 0 & 0 & 0 & 0 & 0 & 0 & 0 & 0 & 0 & 0 & 0 & 0 & 0 & 0 & 0 & 0 & 0 & 0 & 0\\
1 & 0 & 0 & 0 & 0 & 0 & 0 & 0 & 0 & 0 & 0 & 0 & 0 & 0 & 0 & 0 & 0 & 0 & 0 & 0 & 0\\
2 & 0 & 0 & 0 & 0 & 0 & 0 & 0 & 0 & 0 & 0 & 0 & 0 & 0 & 0 & 0 & 0 & 0 & 0 & 0 & 0\\
3 & 0 & 0 & 0 & 0 & 0 & 0 & 0 & 0 & 0 & 0 & 0 & 0 & 0 & 0 & 0 & 0 & 0 & 0 & 0 & 0\\
4 & 0 & 0 & 0 & 0 & 0 & 0 & 0 & 0 & 0 & 0 & 0 & 0 & 0 & 0 & 0 & 0 & 0 & 0 & 0 & 0\\
5 & 0 & 0 & 0 & 0 & 0 & 0 & 0 & 0 & 0 & 0 & 0 & 0 & 0 & 0 & 0 & 0 & 0 & 0 & 0 & 0\\
6 & 0 & 0 & 0 & 0 & 0 & 0 & 0 & 0 & 0 & 0 & 0 & 0 & 0 & 0 & 0 & 0 & 0 & 0 & 0 & 0\\
7 & 0 & 0 & 0 & 0 & 0 & 0 & 0 & 0 & 0 & 0 & 0 & 0 & 0 & 0 & 0 & 0 & 0 & 0 & 0 & 0\\
8 & 0 & 0 & 0 & 0 & 0 & 0 & 0 & 0 & 0 & 0 & 0 & 0 & 0 & 0 & 0 & 0 & 0 & 0 & 0 & 0\\
9 & 0 & 0 & 0 & 0 & 0 & 0 & 0 & 0 & 0 & 0 & 0 & 0 & 0 & 0 & 0 & 0 & 0 & 0 & 0 & 0\\
10 & 0 & 0 & 0 & 0 & 0 & 0 & 0 & 0 & 0 & 0 & 0 & 0 & 0 & 0 & 0 & 0 & 0 & 0 & 0 & 0\\
11 & 0 & 0 & 0 & 0 & 3 & 3 & 3 & 3 & 3 & 3 & 3 & 3 & 3 & 3 & 3 & 3 & 3 & 3 & 3 & 3\\
12 & 0 & 0 & 0 & 0 & 3 & 3 & 3 & 3 & 3 & 3 & 3 & 3 & 3 & 3 & 3 & 3 & 3 & 3 & 3 & 3\\
13 & 0 & 0 & 0 & 0 & 250 & 250 & 250 & 250 & 250 & 250 & 250 & 250 & 250 & 250 & 250 & 250 & 250 & 250 & 250 & 250\\
14 & 0 & 0 & 0 & 0 & 400 & 400 & 400 & 400 & 400 & 400 & 400 & 400 & 400 & 400 & 400 & 400 & 400 & 400 & 400 & 400\\
15 & 0 & 0 & 0 & 0 & 600 & 600 & 600 & 600 & 600 & 600 & 600 & 600 & 600 & 600 & 600 & 600 & 600 & 600 & 600 & 600\\
16 & 0 & 0 & 0 & 0 & 1000 & 1000 & 1000 & 1000 & 1000 & 1000 & 1000 & 1000 & 1000 & 1000 & 1000 & 1000 & 1000 & 1000 & 1000 & 1000\\
17 & 0 & 0 & 0 & 0 & 2000 & 2000 & 2000 & 2000 & 2000 & 2000 & 2000 & 2000 & 2000 & 2000 & 2000 & 2000 & 2000 & 2000 & 2000 & 2000\\
18 & 0 & 0 & 0 & 0 & 4000 & 4000 & 4000 & 4000 & 4000 & 4000 & 4000 & 4000 & 4000 & 4000 & 4000 & 4000 & 4000 & 4000 & 4000 & 4000\\
19 & 0 & 0 & 0 & 0 & 20000 & 20000 & 20000 & 20000 & 20000 & 20000 & 20000 & 20000 & 20000 & 20000 & 20000 & 20000 & 20000 & 20000 & 20000 & 20000\\
20 & 0 & 0 & 0 & 0 & 20000 & 20000 & 20000 & 20000 & 20000 & 20000 & 20000 & 20000 & 20000 & 20000 & 20000 & 20000 & 20000 & 20000 & 20000 & 20000\\
\end{tabular}
\caption{\footnotesize Biasing potential $W(\psi)$ for window 2 of an overstepped walker with tension, with $\psi = (\psi_1, \psi_2, \psi_3)$ a 3-dimensional order parameter. Due to their definitions, states with $\psi_2>0$ and $\psi_3 \neq 3$ are impossible.  \label{recovery_US_OP2_T}}
\end{center}
\end{table*}

 \begin{table*}
\begin{center}
\begin{tabular}{c | c c c c | c c c c c c c c c c c c c c c c}
&\multicolumn{4}{c |}{$\psi_2=0$, $\psi_3=$} & \multicolumn{16}{c}{$\psi_3=3$, $\psi_2=$}\\
$\psi_1$ & 0 & 1 & 2 & 3 & 1 & 2 & 3 & 4 & 5 & 6 & 7 & 8 & 9 & 10 & 11 & 12 & 13 & 14 & 15 & 16 \\
\hline
0 & 0 & 0 & 0 & 0 & 0 & 0 & 0 & 0 & 0 & 0 & 0 & 0 & 0 & 0 & 0 & 0 & 0 & 0 & 0 & 0\\
1 & 0 & 0 & 0 & 0 & 1 & 1 & 1 & 1 & 1 & 1 & 1 & 1 & 1 & 1 & 1 & 1 & 1 & 1 & 1 & 1\\
2 & 0 & 0 & 0 & 0 & 1 & 1 & 1 & 1 & 1 & 1 & 1 & 1 & 1 & 1 & 1 & 1 & 1 & 1 & 1 & 1\\
3 & 0 & 0 & 0 & 0 & 1 & 1 & 1 & 1 & 1 & 1 & 1 & 1 & 1 & 1 & 1 & 1 & 1 & 1 & 1 & 1\\
4 & 0 & 0 & 0 & 0 & 1 & 1 & 1 & 1 & 1 & 1 & 1 & 1 & 1 & 1 & 1 & 1 & 1 & 1 & 1 & 1\\
5 & 0 & 0 & 0 & 0 & 1 & 1 & 1 & 1 & 1 & 1 & 1 & 1 & 1 & 1 & 1 & 1 & 1 & 1 & 1 & 1\\
6 & 0 & 0 & 0 & 0 & 1 & 1 & 1 & 1 & 1 & 1 & 1 & 1 & 1 & 1 & 1 & 1 & 1 & 1 & 1 & 1\\
7 & 0 & 0 & 0 & 0 & 2 & 2 & 2 & 2 & 2 & 2 & 2 & 2 & 2 & 2 & 2 & 2 & 2 & 2 & 2 & 2\\
8 & 0 & 0 & 0 & 0 &  400 & 400 & 400 & 400 & 400 & 400 & 400 & 400 & 400 & 400 & 400 & 400 & 400 & 400 & 400 & 400\\
9 & 0 & 0 & 0 & 0 & 300 & 300 & 300 & 300 & 300 & 300 & 300 & 300 & 300 & 300 & 300 & 300 & 300 & 300 & 300 & 300\\
10 & 0 & 0 & 0 & 0 &  300 & 300 & 300 & 300 & 300 & 300 & 300 & 300 & 300 & 300 & 300 & 300 & 300 & 300 & 300 & 300\\
11 & 0 & 0 & 0 & 0 &  300 & 300 & 300 & 300 & 300 & 300 & 300 & 300 & 300 & 300 & 300 & 300 & 300 & 300 & 300 & 300\\
12 & 0 & 0 & 0 & 0 & 300 & 300 & 300 & 300 & 300 & 300 & 300 & 300 & 300 & 300 & 300 & 300 & 300 & 300 & 300 & 300\\
13 & 0 & 0 & 0 & 0 & 0 & 0 & 0 & 0 & 0 & 0 & 0 & 0 & 0 & 0 & 0 & 0 & 0 & 0 & 0 & 0\\
14 & 0 & 0 & 0 & 0 & 0 & 0 & 0 & 0 & 0 & 0 & 0 & 0 & 0 & 0 & 0 & 0 & 0 & 0 & 0 & 0\\
15 & 0 & 0 & 0 & 0 & 0 & 0 & 0 & 0 & 0 & 0 & 0 & 0 & 0 & 0 & 0 & 0 & 0 & 0 & 0 & 0\\
16 & 0 & 0 & 0 & 0 & 0 & 0 & 0 & 0 & 0 & 0 & 0 & 0 & 0 & 0 & 0 & 0 & 0 & 0 & 0 & 0\\
17 & 0 & 0 & 0 & 0 & 0 & 0 & 0 & 0 & 0 & 0 & 0 & 0 & 0 & 0 & 0 & 0 & 0 & 0 & 0 & 0\\
18 & 0 & 0 & 0 & 0 & 0 & 0 & 0 & 0 & 0 & 0 & 0 & 0 & 0 & 0 & 0 & 0 & 0 & 0 & 0 & 0\\
19 & 0 & 0 & 0 & 0 & 0 & 0 & 0 & 0 & 0 & 0 & 0 & 0 & 0 & 0 & 0 & 0 & 0 & 0 & 0 & 0\\
20 & 0 & 0 & 0 & 0 & 0 & 0 & 0 & 0 & 0 & 0 & 0 & 0 & 0 & 0 & 0 & 0 & 0 & 0 & 0 & 0\\
\end{tabular}
\caption{\footnotesize Biasing potential $W(\psi)$ for window 3 of an overstepped walker with tension, with $\psi = (\psi_1, \psi_2, \psi_3)$ a 3-dimensional order parameter. Due to their definitions, states with $\psi_2>0$ and $\psi_3 \neq 3$ are impossible.  \label{recovery_US_OP3_T}}
\end{center}
\end{table*}

\subsection{Kinetic simulations of blunt-ended displacement}
\label{protocols_disp}
I performed 10 completely independent implementations of FFS to estimate displacement rate for zero-base toeholds in oxDNA. In each case, I measured the flux from the initial state (incumbent/substrate duplex, separate invader) to the final state (invader/substrate duplex, separate incumbent) using the order parameter outlined in Table \ref{blunt_end_FFS_op}. The order parameter involves the number of {\em nearly formed} base pairs; a base pair is defined as nearly formed if and only if the following criteria hold.
\begin{itemize}
\item The separation of hydrogen bonding sites is $\leq 0.85$\,nm.
\item The hydrogen-bonding potential consists of a separation dependent factor multiplied by a number of modulating angular factors. At most one of these factors that contributes multiplicatively to the hydrogen-bonding energy is zero.  
\end{itemize}
Physically, these conditions mean that the bases are close and fairly well aligned, but not forming a strong base pair.

In Ref. \onlinecite{Srinivas2013}, it was observed that all successful instances of blunt-ended displacement began with attachment at one of the three bases at either end of the displacement domain. In this work, I use an order parameter that assumes a pathway in which displacement begins with attachment at either end of the displacement domain. This enables more efficient sampling of this dominant pathway at the cost of less efficient sampling of alternatives, which are assumed to be irrelevant.

For each of the 10 independent FFS implementations, I ran 10 simulations to measure the initial flux. In each case, 100 states at the interface $\lambda_{-1}^0$ were collected, giving 10000 in total.  All simulations were initialised in the same state, but then separately thermalised before flux measurement.

For each subsequent interface, a large number of trajectories were launched and those that reached the next interface before returning to $Q=-2$ were saved. The number of attempts made at each interface are given in Table~\ref{blunt_end_FFS_protocols}. 

\subsection{Kinetic simulations of walker recovery}
The procedure for kinetic simulations of walker recovery by fuel invasion was generally similar to that outlined in Section \ref{protocols_disp} for blunt-ended displacement. Order parameter definitions and simulation details are presented in Tables \ref{recovery_FFS_OP} and \ref{recovery_FFS_protocols} (note that, in this case, I make no assumptions about where initial attachment occurs).

In simulations, the effective concentration of the walker and fuel are much higher than typical of experiment. As a result, the ratio of time spent whilst displacement is in progress to time spent waiting for fuel-binding is much higher. It is possible, therefore, that in our system recovery of the foot is not an effectively instantaneous process compared to the lifetime of the overstepped state, as would be expected in experiment. To avoid this confusion, I did not include time spent with the fuel partially bound to the walker in our estimate of the initial flux -- a simple way to do this was to restart flux simulations if they reached the interface $\lambda_6^7$ in Table \ref{recovery_FFS_OP}  (the point at which the fuel starts to form base pairs with the walker), rather than waiting until $\lambda_{10}^{11}$ before doing so. The overall procedure is in fact equivalent to measuring the flux from $Q=-2$ to $Q=7$, and then the subsequent probability of going on from $Q=7$ to $Q=11$ rather than back to $Q=-2$.

The fuel is only 16 bases long, and hence cannot fully displace the track from the foot -- the final six base pairs must break spontaneously. In this work I use the rate at which this maximally displaced state is reached as a proxy for the rate of recovery (foot-lifting). From the known behaviour of toehold-mediated strand displacement, one would expect detachment of this region of six base pairs to be faster than reverse-displacement of the fuel by the track, a process that would be even slower than normal due to the  need to overcome two mismatches.\cite{Srinivas2013,Machinek2014} Consequently the probability of recovery once the six-base-pair state is reached should be high, justifying the approximation in this work.

\subsection{Thermodynamic simulations of walker recovery}
I performed thermodynamic simulations of walker recovery using VMMC aided by US and WHAM. I divided state space into three overlapping `windows', and ran 10 simulations confined to each window. The order parameter $\psi({\bf r}^N, {\bf q}^N)$ used to define the umbrella potential was three-dimensional:
\begin{itemize}
\item $\psi_1$: the number of  base pairs between the overstepped foot and the track with a base-paring interaction more negative than $-0.596$\,kcal\,mol$^{-1}$.
\item $\psi_2$: the number of  base pairs between the overstepped foot and the fuel with a base-paring interaction more negative than $-0.596$\,kcal\,mol$^{-1}$.
\item $\psi_3$: a measure of the proximity of the fuel and the overstepped walker foot, measured as the minimum distance $d$ between hydrogen-bonding sites of bases in the fuel and foot domain to which it binds. For historical reasons, in the simulations without tension this distance $d$ was only calculated using base pairs that can form in the intended foot-track duplex; in the simulations with fuel, it was calculated using any pair of bases in the two domains. Although this leads to different sampling behaviour, it does not invalidate the schemes. 4 states of $\psi_3$ were used:
\begin{itemize}
\item $\psi_3=0$: $d>8.52$\,nm.
\item $\psi_3=1$: $8.52$\,nm $ \geq d>4.26$\,nm.
\item $\psi_3=2$: $4.26$\,nm $ \geq d>2.56$\,nm.
\item $\psi_3=3$: $2.56$\,nm $\geq d$.
\end{itemize}
\end{itemize}

The biasing weights used for each simulation window are outlined in Tables \ref{recovery_US_OP1_noT}, \ref{recovery_US_OP2_noT}, \ref{recovery_US_OP3_noT}, \ref{recovery_US_OP1_T}, \ref{recovery_US_OP2_T} and \ref{recovery_US_OP3_T}. For a given window, systems were initialised in the same state and then thermalised prior to data collection. $10^9$ and $10^{10}$ VMMC steps were used for  initialisation and data collection for the first window in each case; $2 \times10^9$ and $2 \times 10^{10}$ were used for the second and third windows for both systems.  Data from distinct windows were combined using the WHAM algorithm.\cite{Kumar1992} 

\section{Results and error analysis}
\subsection{Kinetic simulations of blunt-ended displacement}
The results for the 10 independent FFS implementations are presented in Table \ref{blunt_end_FFS_protocols}. As is evident, the random error is substantial (and much larger than would be estimated from the simple argument outlined in Section \ref{methods:LD}). Averaging the results and taking a standard error on the mean would lead to an estimate of $0.027\pm 0.013$\,s$^{-1}$, significantly higher than reported in Ref. \onlinecite{Srinivas2013} ($0.005$\,s$^{-1}$). The new result, however, is not less consistent with the experimental data. \teo{The previous result suggested that zero-base toehold was slower relative to a six-base toehold in oxDNA than in reality}. \teo{The new measurements also have} no consequences for the physical interpretation provided in Ref. \onlinecite{Srinivas2013}, which was not based on the behaviour of the zero-base toehold itself but on observations related to why longer toeholds (which were sampled more reliably) had a high probability of detaching before displacement could proceed.

\subsection{Kinetic simulations of walker recovery}
The results for the 10 independent FFS implementations of walker recovery with and without tension applied to the track are presented in Table \ref{recovery_FFS_protocols}. Once again, the random errors are substantial due to the difficulty of measuring such slow processes with large systems, but the overall difference between systems with and without tension is clear.

\subsection{Thermodynamic simulations of walker recovery}
\begin{figure}
\centering
\includegraphics[width=0.35\textwidth,angle=-90]{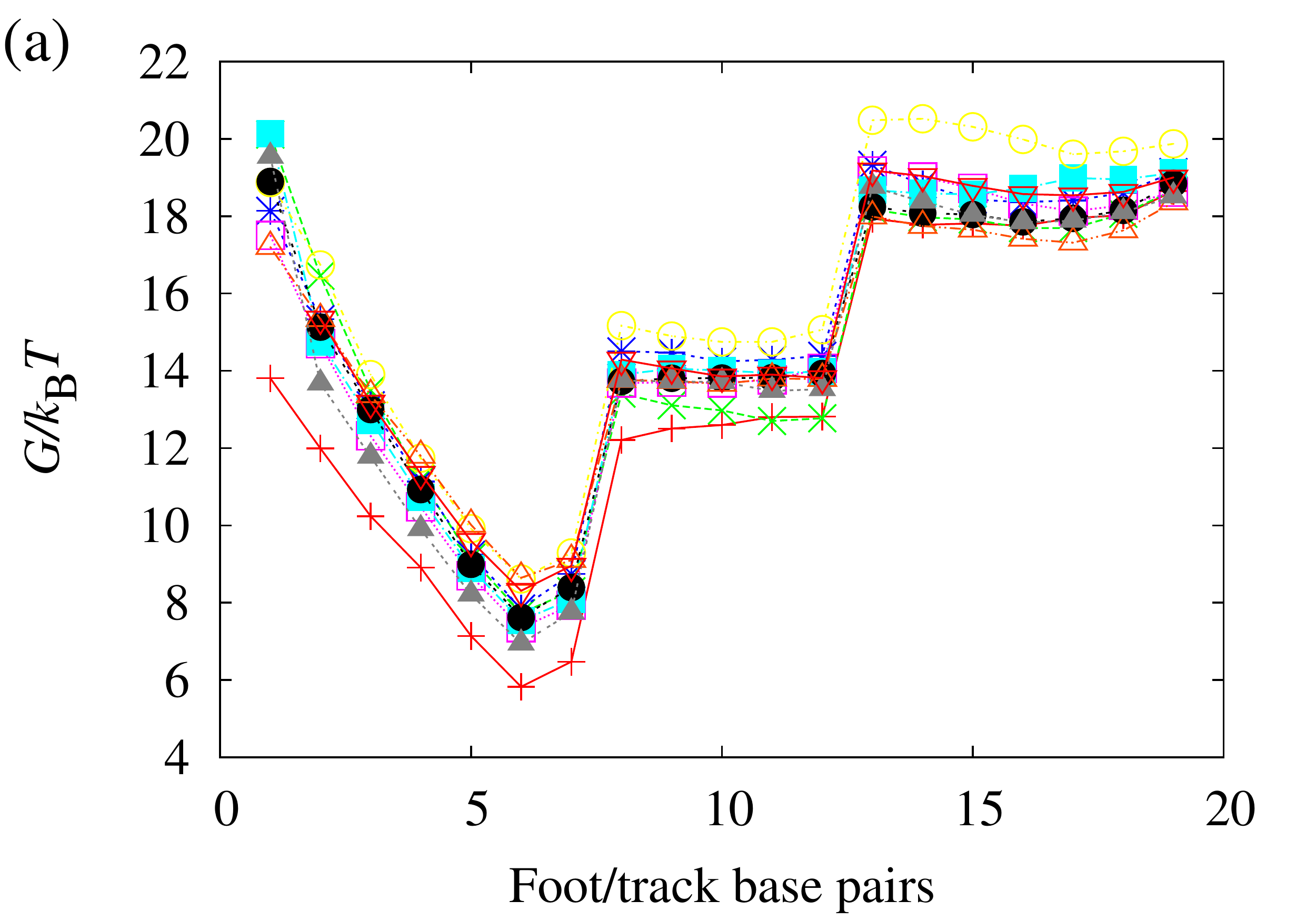}\\
\includegraphics[width=0.35\textwidth,angle=-90]{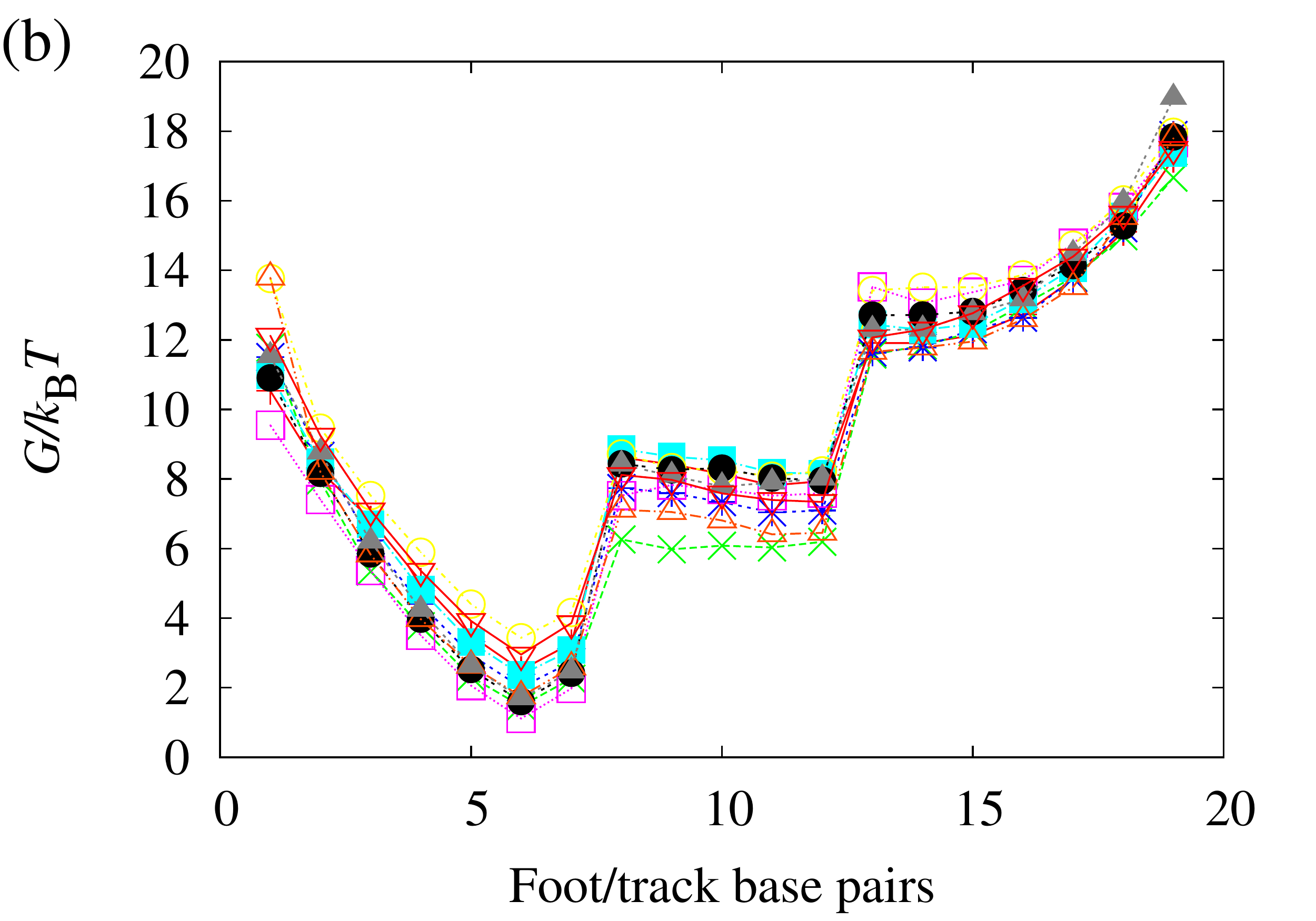}
\centering \caption{\footnotesize Results of 10 independent estimates of the free-energy profile ($G$) for fuel displacement of the track; (a) without tension and (b) with the track under 14.6\,pN of tension. The combination of the ten separate estimates into a single profile for each case is shown in Fig. 10 of the main text. Free energies are given as a function of the number of foot/track base pairs ($\psi_1$), subject to the constraint that the fuel is bound to the foot by at least one base pair ($\psi_2>0$, $\psi_3=3$). Free energies are reported relative a system without bound fuel  ($\psi_2=0$, $\psi_3 = 0,1,2,3$).
\label{fig_VMMC_noise}}
\end{figure}

The overall thermodynamic data for simulations of walker recovery are presented in the main text. The random error can be estimated by grouping the data into 10 independent sets (each one containing one of the independent simulations from each window), and running the WHAM algorithm on each one separately. The results of such a procedure are shown in Fig. \ref{fig_VMMC_noise} -- the standard error on the mean for the free-energy relative to the unbound state is typically $\sim 0.2-0.3$\,$k_{\rm B}T$.

\end{appendix}

\end{document}